\documentstyle[aas2pp4]{article} 
\newcommand\kms{km~s$^{-1}$}
\newcommand\etal{{\it et al.}}
\newcommand\vsun{V$_\odot$}
\newcommand\msun{M$_\odot$}
\newcommand\HI{\ion{H}{1}}
\newcommand\NII{[\ion{N}{2}]}

\newcommand\Ha{H$\alpha$}
\newcommand\MgIb{\ion{Mg}{1} b}
\newcommand\hrs{$^{h}$}
\newcommand\mts{$^{m}$}
\newcommand\magarc{{\rm mag\,arcsec$^{-2}$}}
\newcommand\email{\affil}

\begin{document}
\title{Kinematic Evidence of Minor Mergers in Normal Sa Galaxies:\\
NGC~3626, NGC~3900, NGC~4772 and NGC~5854}

\author{Martha P. Haynes\altaffilmark{1,2}}  
\affil{Center for Radiophysics and Space
Research and National Astronomy and Ionosphere Center\altaffilmark{3},
Cornell University, Space Sciences Building,
Ithaca NY 14853}
\email{haynes@astrosun.tn.cornell.edu}

\author{Katherine P. Jore\altaffilmark{1,2}} 
\affil{Department of Physics and Astronomy, University of Wisconsin
at Stevens Point, Stevens Point, WI 54481}
\email{kjore@uwsp.edu}

\author{Elizabeth A. Barrett}
\affil{Department of Astronomy and
Center for Radiophysics and Space
Research, Cornell University, Space Sciences Building,
Ithaca NY 14853}
\email{eab15@cornell.edu}

\author{Adrick H. Broeils\altaffilmark{1,4}}
\affil{Center for Radiophysics and Space
Research, Cornell University, Space Sciences Building,
Ithaca NY 14853}
\email{adrick.broeils@nl.origin-it.com}

\author{Brian M. Murray}
\affil{Linfield College, 900 SE Baker St., McMinnville, OR 97128}
\email{bmurray@linfield.edu}

\altaffiltext{1}{Visiting Astronomer, National Radio Astronomy Observatory
which is a facility of the National Science Foundation operated under
a cooperative agreement by Associated Universities, Inc.}

\altaffiltext{2}{Visiting Astronomer, Kitt Peak National Observatory
which is part of the National Optical Astronomy Observatories,
operated by the Association of Universities for Research in Astronomy,
Inc. (AURA) under a cooperative agreement with the National Science
Foundation.}  

\altaffiltext{3}{The National Astronomy and Ionosphere Center is
operated by Cornell University under a cooperative agreement with the
National Science Foundation.}

\altaffiltext{4}{Now at Origin Nederland B.V., Nieuwegein, The Netherlands.}

\begin{abstract}
BVRI and \Ha ~imaging and long--slit optical spectroscopic
data are presented for four morphologically normal and relatively 
isolated Sa galaxies, NGC~3626, NGC~3900, NGC~4772 and NGC~5854. 
VLA \HI ~synthesis imaging is presented for the first three objects.
In all four galaxies, evidence of kinematic decoupling of ionized
gas components is found in the long--slit spectroscopic data;
the degree and circumstances of the distinct kinematics vary
from complete counterrotation of all of the gas from all of
the stars (NGC~3626) to nuclear gas disks decoupled
from the stars (NGC~5854) to anomalous velocity central gas components
(NGC~3900 and NGC~4772). In the three objects mapped in \HI,
the neutral gas extends far beyond the optical radius, $R_{HI}/R_{25}$ $\ge$ 2.
In general, the \HI ~surface density is very low and the outer \HI ~is patchy
and asymmetric (NGC~3900) or found in a distinct ring, exterior to
the optical edge (NGC~3626 and NGC~4772). While the overall \HI
~velocity fields are dominated by circular motions, strong warps are
suggested in the outer regions by bending of the minor axis isovelocity
contours (NGC~3900) and/or systematic shifts in position angle between inner and
outer rings (NGC~3626 and NGC~4772). In the interior, coincidence is found
between the \Ha ~and \HI ~in rings, sometimes partial and
crisscrossed by dustlanes. Optical imaging is also presented
for NGC~4138 previously reported by Jore \etal ~(1996) to 
show counterrotating stellar components. The multiwavelength
evidence is interpreted in terms of the kinematic ``memory''
of past minor mergers in objects that otherwise exhibit
no morphological signs of interaction.


\end{abstract}

\section{Introduction}

While mergers between comparable mass galaxies may be responsible for
some of the most dramatic extragalactic events, minor mergers with small
mass satellites may play a subtle but nonetheless critical role in the
evolution of garden--variety spirals, including perhaps the Milky Way.
It is easy to recognize the fireworks of the former by the tides, starbursts
and nuclear activity they trigger. The latter however may lurk hidden
to morphological inspection. Minor mergers are suggested to produce
the X--shaped structures seen in some peculiar S0's (Mihos \etal 
~1995), the departures from axisymmetry seen in a significant fraction
of spiral disks (Zaritsky \& Rix 1997), the driving mechanism for strong
stellar bars (Laine \& Heller 1999) and the peculiar extended 
counter--rotating disks
seen in some early--type spirals, e.g. NGC~7217 (Merrifield \& Kuijken 1994;
Buta \etal ~1995),
NGC~3593 (Bertola \etal ~1996; Corsini \etal ~1998), 
NGC~3626 (Ciri \etal ~1995) and NGC~4138 (Jore \etal ~1996). 

Satellites are commonly found in the vicinity of normal galaxies (Zaritsky \etal
~1997). If the Local Group is at all typical of spiral extragalactic environments,
then the characteristics of such satellites are familiar. Interaction of the Milky
Way with its low mass satellites is hypothesized to explain a variety of
local kinematic peculiarities: the thick disk, the outer flare and warp, the Magellanic
Stream and the high velocity \HI ~clouds. 

Several authors have begun to address numerically the complex phenomenon of
minor mergers of gas--rich satellites with more massive spiral companions.
An immediate worry is how to avoid over--heating the inner disk of the primary
or disrupting it entirely. Minor mergers of low--density, small mass, gas--rich
companions may avoid vertical heating, thereby preserving the victim disk. 
Predictions of the
observable consequences of the accretion of a small mass companion by a spiral
include the warping or actual spreading out of the disk
(Quinn \etal ~1993).  As the satellite material sinks towards the center
of the large galaxy, the gas, unlike the stars, may lose angular momentum
because of dissipation (Hernquist \& Mihos 1995). The resultant inflow of
gas towards the center may trigger a nuclear starburst (Mihos \& Hernquist
1994) or fuel nuclear activity (Taniguchi 1999).
Because of the predicted heating, even a satellite with one tenth the mass
will thicken the disk and possibly build--up the bulge, driving the morphology
of the post merger primary towards an earlier spiral type. The discovery of
kinematically decoupled disks in early spirals may thus prove the efficacy
of minor merger models.

Intrinsic galaxy properties arise from two principal components: absolute
``scale'' (size, luminosity, mass) and ``form'' (morphology, bulge--to--disk ratio, color, 
gas content, star formation rate). It is well-known (e.g. Roberts \& Haynes 1994) that the 
early spiral types show a much greater spread in their characteristic properties 
related to ``form'' than do their later--type counterparts.
As a part of a study to investigate the heterogeneity and dark matter content
of the Sa galaxy class, we have investigated the kinematics of gas and stars
in a sample of morphologically normal, isolated galaxies classified as
Sa in the {\it Revised Shapley Ames Catalog} (Sandage \& Tammann 1987; RSA).
In the course of that study, we have obtained major and minor axis long slit 
spectra for both stars and gas for a sample of 20 objects; for nine of them, 
we have also obtained \HI ~synthesis maps yielding also
two--dimensional \HI ~velocity fields.
A summary of results focussing on the data presentation and mass modelling
was presented in Jore (1997). The analysis of the mass modelling and
stellar velocities will be presented elsewhere (Jore \etal, in preparation).

During the course of that work, a number of
unusual cases of kinematic decoupling were discovered. In Jore \etal
~(1996), we discussed the distinctive case of NGC~4138, an isolated Sa
with two extended coplanar counter--rotating stellar disks embedded in a huge
HI disk that co--rotates with the secondary stellar component. In a separate
paper (Jore \etal ~2000), we discuss five Sa galaxies, NGC~1169, 3623, 4866, 5377 and 5448,
which represent a range of environments and morphologies within the Sa class.
Of those objects, NGC~1169 and NGC~5448, exhibit no peculiar optical kinematics
but their outer \HI ~disks are strongly warped and kinematically skewed.
NGC~3623, whose companions in the Leo Triplet are known to
be interacting, exhibits a truncated \HI ~disk and central gas decoupling.
NGC~4866, viewed nearly edge--on, and NGC~5377 show doubling of their central
gas components, as well as outer \HI ~warps. Here, we present four 
additional examples: NGC 3626, NGC~3900,
NGC~4772 and NGC~5854. Along with NGC~4138, each of these morphologically 
normal Sa galaxies exhibits peculiar kinematics that betray a disturbed past.
We propose that each represents a different stage and circumstance of 
involvement in a relatively minor merger event with a
gas--rich companion that is no longer recognizable.

This paper presents evidence of this past interaction through a
combination of optical imaging and spectroscopy and \HI ~synthesis
mapping. In Section 2, we discuss the acquisition and reduction of the
body of data: broad-- and narrow--band optical images, optical long--slit spectra,
and \HI ~synthesis maps. The kinematic evidence of past interaction is
presented for each galaxy in Section 3. Section 4 gives a discussion of 
the observational results in the context of the minor merger scenario.

\section{Observations}

The original observing program was designed to provide the
details of the galaxy dynamics for the purpose of modelling the dark
matter distribution in Sa galaxies and, as such,
consisted of long--slit major and minor axis
spectra of both stars and ionized gas. For a subset, principally
gas--rich objects showing evidence in the optical spectra 
of kinematic peculiarities,
\HI ~synthesis mapping and photometric broad-- and narrow-- band
imaging was also undertaken.

The basic properties of these four galaxies plus
NGC~4138 discussed by Jore \etal ~(1996) are summarized
in Table 1. The first entries include the (B1950)
R.A. and Dec. and the morphological classifications as given in both the
RSA (Sandage \& Tammann 1987) and the {\it Third Reference Catalog of Bright Galaxies}
(de Vaucouleurs \etal ~1991; RC3). A morphological description of the \Ha ~emission
is then given, followed by a
set of radial measures: $D_{25} \times d_{25}$, in arcminutes,
from the RC3 and the corresponding radius R$_{25}$, in arcseconds; 
R$_{B25}$, in arcseconds, derived from the $B$ image;
R$_{I83L}$, in arcseconds, the radius encompassing 83\% of the total $I$ band light,
following Haynes \etal ~(1999a); and
R$_{orc}$, in arcseconds, the maximum radius at which the optical rotation curve
is derived, for any species, stars or gas.
The corrected integrated blue magnitude, B$_T^0$, and the color indices,
$(U-B)$ and $(B-V)$, all from the RC3, are then listed, along with the optical
heliocentric velocity, V$_\odot$, in \kms, and the adopted distance, in Mpc. 
Since all of these galaxies are nearby, distances are adopted from available
estimates accounting for group membership and local peculiar motions and/or 
supernova distance moduli. The total $B$ luminosity, L$_B$,
is derived from B$_T^0$ and the distance. The total apparent magnitudes 
m$_B$, m$_V$, m$_R$, and m$_I$, and their estimated errors, have been derived 
from the images following Haynes \etal ~(1999a). Likewise, the ellipticity $\epsilon = 
(1 - {b \over a})$ and position angle are the mean disk properties extracted
from the $I$ band isophotal fits using the ``marking the disk'' method of 
Haynes \etal ~(1999a). Finally, the adopted $R$ band bulge--to--disk ratio
B/D$_R$ is given.

\begin{deluxetable}{lccccc}
\tablewidth{0pt}
\tablenum{1}
\tablecaption{Galaxy Properties \label{tabprop}}
\tablehead{
\colhead{} &  \colhead{N~3626} & \colhead{N~3900} & \colhead{N~4138} & \colhead{N~4772} & \colhead{N~5854} 
\\[.2ex]
\colhead{} &  
\colhead{U~6343} & \colhead{U~6786} & \colhead{U~7139} & \colhead{U~8021} & \colhead{U~9726}}
\startdata
R.A.(B1950)        & 111725.9 & 114633.3 &  120659.3 &  125055.9 & 150516.2 \nl
Dec.(B1950)        & +183756  &  +271806 &   +435757 &   +022627 & +024537 \nl  
RSA Type\tablenotemark{a} & Sa       & S(r)a    & S(r)a pec & Sa:      & Sa \nl
RC3 Type\tablenotemark{b} & RLAT+..  & .LAR+..  & .LAR+..   & .SAS1..  & LBS+./\nl 
H$\alpha$ morph \tablenotemark{c} & n+cr & (trace) & n+cb+or & n+cb+or & \nodata\nl
\nl
$D_{25} \times d_{25}$\tablenotemark{b}   & 2.69x 1.95 & 3.16 x 1.70 & 2.57 x 1.70 & 3.39 x 1.70 & 2.75 x 0.79\nl
R$_{25}$ (\arcsec)\tablenotemark{b} & ~81 & ~96 & ~78 & 102 & ~84\nl
R$_{B25}$ (\arcsec) & ~76 & ~97 & ~85 & 132 & ~78\nl
R$_{I83L}$ (\arcsec) & ~56 & ~75 & ~54 & 120 & ~48 \nl
R$_{orc}$ (\arcsec)\tablenotemark{d} & ~45 & ~53 & ~50 & ~58 & ~51 \nl
\nl
B$_T^0$\tablenotemark{b} & 11.75    & 12.24    & 12.14 & 11.90    & 12.68  \nl
$(U-B)$\tablenotemark{b}   & 0.30 & \nodata & 0.31  & 0.26 & 0.29\nl
$(B-V)$\tablenotemark{b}   & 0.81 & 0.82    & 0.83  & 0.83 & 0.81\nl
V$_\odot$ (\kms)\tablenotemark{b} & 1493 & 1798 & ~888 & 1040 & 1663\nl
Distance (Mpc) & ~~24 & ~~27 & ~~16 & ~~16 & ~~24\nl
L$_B$ ($10^{10}$ $L_{\odot}$) & 1.8 & 1.4 & 0.6 & 0.7 & 0.8\nl
\nl
m$_B$  & 11.91 $\pm$.04 & 12.40 $\pm$.02 & 12.12 $\pm$.04 & 11.97 $\pm$.04 & \nodata\nl
m$_V$  & 11.11 $\pm$.02 & 11.65 $\pm$.02 & 11.31 $\pm$.02 & 11.15 $\pm$.03 & 11.20 $\pm$.10\nl
m$_R$  & 10.59 $\pm$.02 & 11.08 $\pm$.02 & 10.73 $\pm$.03 & 10.58 $\pm$.02 & 10.64 $\pm$.02\nl
m$_I$  & \nodata  & 10.42$\pm$.02 & 10.03$\pm$.02 & \nodata & \nodata\nl 
\nl
$\epsilon$        & 0.32$\pm$0.02 & 0.48$\pm$0.02 & 0.39$\pm$0.02 & 0.48$\pm$0.03 & 0.67$\pm$0.02 \nl
P.A. ($^\circ$)   & 164$\pm$3 & ~~4$\pm$1 & 150$\pm$1 & 148$\pm$2 & ~55$\pm$1 \nl 
B/D$_R$ & 0.37 & 0.27 & 0.19 & 0.44 & 0.82\nl
\enddata
\tablenotetext{a}{From RSA.}
\tablenotetext{b}{From RC3.}
\tablenotetext{c}{Code for morphology: n=nuclear; cr=core ring; cb=elongated core; or=outer ring.}
\tablenotetext{d}{Maximum extent of optical rotation curve traced by any species.}
\end{deluxetable}

The acquisition and reduction of the
multiwavelength data sets are discussed here separately.

\subsection{Optical Images}

$BVR$ and \Ha ~images were obtained 
with the KPNO 0.9m telescope in its f/7.5 configuration over 
the nights of April 19--23, 1996. Additionally,
$I$ band images, some taken under non-photometric conditions, were obtained
with the same instrument by R. Giovanelli and MPH as part of another project.
For all images, the 2048$\times$2048 pixel T2KA CCD was used. 
Its read noise is 4 e$^-$ and the gain was set to 5.4 e$^-$ ADU$^{-1}$, except
for the the $I$ band images, when the gain was set to 3.6 e$^-$ ADU$^{-1}$. 
With the field corrector in, the images had a pixel scale of 0.68\arcsec.
In this section, we describe the image acquisition, processing and
analysis techniques.

\subsubsection{Broadband BVRI Images}

The broadband images were exposed over the whole chip, giving 
a total field of view of 23\arcmin ~on a side.
The images were trimmed, bias subtracted, and flat fielded using CCDPROC
in IRAF \footnote{IRAF is distributed by the National Optical Astronomy
Observatories.}.  Flat fields for $B$, $V$ and $R$ were obtained from twilight
sky observations; for the $I$ band, a flat field was constructed from the median
of 30--50 object frames taken throughout the night. Further image reductions made use of the
GALPHOT\footnote{The GALPHOT surface photometry package is a collection of
IRAF/STSDAS scripts originally developed by Wolfram Freudling and John Salzer;
the Cornell version has been further modified and is maintained by MPH.}
package of IRAF/STSDAS scripts for galaxy surface photometry (see Haynes
\etal ~1999a).  Regions of
the image free of galaxy emission and bright stars were interactively
marked. An average value of the sky contribution in these boxes
after masking of the stars within them was subtracted as the
sky contribution. Images were also edited interactively
to remove stars and cosmic rays and convolved to guarantee the same resolution
before combining. Images with three or more observations on the
same filter were scaled and median combined to remove cosmic rays.  For
images with fewer exposures cosmic rays were removed before combining
using the task COSMIC in IRAF and interactively editing all remaining cosmic
rays by hand. The filters, total exposure times, and the seeing values for the
combined images are presented in Table 2 for each galaxy.  

\begin{deluxetable}{lccrcc}
\small
\tablewidth{0pt}
\tablenum{2}
\tablecaption{Details of Imaging Observations\label{tabbvr}}
\tablehead{
\colhead{Object} & \colhead{Filter} & \colhead{Night} & 
\colhead{T$_{exp}$} & \colhead{Seeing}  & \colhead{Note}\\
\colhead{} & \colhead{} & \colhead{} &
\colhead{(sec)} & \colhead{$^{\prime\prime}$} & \colhead{}}
\startdata
N3626 & B & 1 & 3600 & 2.69 & a\nl
  & V & 1 & 1350 & 2.40 & a\nl
  & R & 1 & 1080 & 1.97 \nl
  & I &   & 600  & 2.99 & b\nl
  & ON & 4 & 3600 & 1.98 & b\nl
  & OFF & 4 & 1800 & 1.98 & b\nl
N3900 & B & 3 & 3600 & 2.03 \nl
  & V & 3 & 1350 & 2.02\nl
  & R & 3 & 900 & 2.16 \nl
  & I &   & 600 & 1.62 \nl
  & ON & 5 & 3600 & 1.81 & b\nl
  & OFF & 4 & 1800 & 1.77 & b\nl
N4138 & B & 1 & 3600 & 2.24 \nl
  & V & 1 & 1350 & 1.86 \nl
  & R & 1 & 900 & 1.81 \nl
  & I &   & 600 & 2.12 \nl
  & ON & 5 & 3600 & 2.00 & b\nl
  & OFF & 4 & 1800 & 2.09 & b\nl
N4772 & B & 1 & 3600 & 1.57 \nl
  & V & 3 & 1350 & 1.65 \nl
  & R & 3 & 900 & 1.65 \nl
  & I &   & 600 & 1.92 & b\nl
  & ON & 5 & 3600 & 1.56 & b\nl
  & OFF & 5 & 1800 & 1.64 & b\nl
N5854 & B & 1 & 3600 & 2.50 & b\nl
  & V & 1 & 1350 & 2.19\nl
  & R & 1 & 900 & 1.99\nl
  & I &   & 600 & 2.58 & b\nl
\enddata
\tablenotetext{a}{Tracking problems; image quality poor}
\tablenotetext{b}{Nonphotometric}
\end{deluxetable}

Elliptical isophotal fitting routines were run on the
star edited $BVRI$ images of each galaxy.  These routines require the user to
mark an initial major and minor axis.  The program then fits isophotal
ellipses moving outwards and inwards from the marked initial location and
integrates the flux within the ellipses.  Using these results, disk
parameters such as position angle, ellipticity, and slope are fit and
extrapolated magnitudes are found. Selected values of these
parameters are included in Table 1.

Conversion from instrumental to apparent magnitudes was accomplished
using frequent observations of standard stars (Landolt 1992)
observed throughout each night.  The photometric coefficients for two photometric
nights during which the $BVR$ images were obtained (nights 1 and 
3) are listed in Table 3.  Photometric
coefficients were also determined for night 2; however, they fail to 
reproduce Landolt standard magnitudes accurately and were therefore discarded as
non--photometric.  All images from night 2 were subsequently scaled to 
ones obtained on nights 1
or 3 in order to insure photometric accuracy.

\begin{deluxetable}{lcccc}
\small
\tablewidth{0pt}
\tablenum{3}
\tablecaption{Photometric Solutions for BVR Imaging \label{tabphot}}
\tablehead{
\colhead{Night} & 
\colhead{Filter} &
\colhead{Zeropoint} &
\colhead{Color Coef.} & 
\colhead{Extinc. Coef.} \\
\colhead{} &
\colhead{} &
\colhead{(mag)} &
\colhead{} &
\colhead{}
}
\startdata
1 & B & 21.005 (0.039) & 0.099 (0.019) & 0.237 (0.004) \nl
  & V & 21.311 (0.018) & 0.000 (0.009) & 0.157 (0.004) \nl
  & R & 21.411 (0.013) & 0.009 (0.014) & 0.104 (0.003) \nl
\cr
3 & B & 21.062 (0.007) & 0.075 (0.012) & 0.233 (0.004) \nl
  & V & 21.354 (0.013) & -0.017 (0.010) & 0.155 (0.003) \nl
  & R & 21.424 (0.010) & 0.019 (0.016) & 0.102 (0.003) \nl
\enddata
\end{deluxetable}

Color plots of each galaxy in $B-V$, $B-R$ and $V-R$ were
produced by first aligning all the bandpasses and convolving them to the same
resolution (the poorest of the set, as given in Table 2). The images were then
converted into instrumental magnitudes and
rescaled to apparent magnitudes using the appropriate photometric coefficients.
Final color images were made by subtracting the resultant images and correcting
for galactic extinction.  Values for galactic extinction were calculated using
A$_g$(B) values from the RC3 and the conversions
to other bands given in Rieke \& Lebofsky (1985).
Additionally, model color images were also created from the broadband isophotal
ellipse fitting routines mentioned above.  Using the results of the ellipse
fitting, model broadband images were created, converted into apparent magnitudes
as above, and subtracted to form noiseless model $B-V$, $B-R$ and $V-R$ images.

The separation of bulge and disk components was accomplished
using FITMAP, a two-dimensional decomposition routine written and kindly provided 
by G. Moriondo (Moriondo \etal ~1998).  FITMAP fits a model 
with a specified bulge and exponential disk to the data, adjusting the allowed 
parameters until the optimum model is found as judged by the reduced $\chi^2$ value. 
The bulge shape index $n$ for the generalized exponential law
$I_b(R) = I_{e}exp[-(R/R_e)^{1/n}]$ can be specified as an integer value, with 
$n=1$, an exponential bulge and $n=4$, a de Vaucouleurs bulge. 
In order to speed processing
time, the fit was performed only on an averaged quadrant of the galaxy.  In this
data set, the average was typically a combination of all four quadrants,
achieved by folding the image along both major and minor axes, except where 
bright stars were found close to the galaxy center. All filters were fitted
separately to allow comparison of the results obtained with different bandpasses.

\begin{deluxetable}{lcccccccll}
\small
\tablewidth{0pt}
\tablenum{4}
\tablecaption{Results of 2D Bulge--Disk Decomposition.\label{tabbdrat} }
\tablehead{
\colhead{} &
\colhead{} &
\multicolumn{2}{c}{Exponential Disk} &
\multicolumn{2}{c}{$n=2$ Bulge} &
\colhead{} &
\colhead{} & 
\colhead{}  \\
\colhead{Object} &
\colhead{Filter} &
\colhead{$\mu_{\circ}$} &
\colhead{$h$} &
\colhead{$\mu_e$} &
\colhead{$R_e$} &
\colhead{$\chi^2$} & 
\colhead{B/D} & \colhead{}\\
\colhead{} &
\colhead{} &
\colhead{($mag/arcsec^2$)} &
\colhead{(arcsec)} &
\colhead{($mag/arcsec^2$)} &
\colhead{(arcsec)} &
\colhead{} & 
\colhead{} &
\colhead{} & \colhead{}
}
\startdata
N3626 & B & 20.94(0.01) & 21.3(0.16) & 18.94(0.05) & 3.2(0.08) & 2.73 & 0.37 \nl
      & V & 20.10(0.01) & 21.1(0.12) & 17.99(0.05) & 3.0(0.06) & 3.75 & 0.37 \nl
      & R & 19.55(0.01) & 20.7(0.13) & 17.50(0.04) & 3.0(0.06) & 3.20 & 0.37 \nl
      & I & \nodata     & 21.3(0.34) & \nodata     & 3.4(0.05) & 1.16 & 0.30 & a\nl
N3900 & B & 21.68(0.02) & 24.4(0.21) & 20.69(0.05) & 4.8(0.18) & 1.55 & 0.25 \nl
      & V & 20.78(0.01) & 23.0(0.16) & 19.77(0.04) & 4.9(0.13) & 1.61 & 0.30 \nl
      & R & 20.15(0.01) & 22.2(0.14) & 19.16(0.04) & 4.8(0.12) & 1.84 & 0.27 \nl
      & I & 19.27(0.01) & 20.7(0.11) & 18.25(0.02) & 4.0(0.06) & 0.79 & 0.25 \nl
N4138 & B & 20.46(0.03) & 16.6(0.14) & 21.44(0.08) & 7.0(0.48) & 2.19 & 0.19 \nl
      & V & 19.67(0.01) & 16.9(0.08) & 19.90(0.04) & 5.2(0.16) & 1.64 & 0.20 \nl
      & R & 19.06(0.01) & 16.7(0.06) & 19.01(0.04) & 4.4(0.11) & 1.72 & 0.19 \nl
      & I & 18.37(0.01) & 17.0(0.05) & 17.95(0.02) & 3.7(0.06) & 1.02 & 0.18 \nl
N4772 & B & 22.81(0.01) & 50.4(0.36) & 21.22(0.01) & 9.6(0.10) & 0.44 & 0.41 \nl
      & V & 21.96(0.01) & 48.4(0.25) & 20.27(0.01) & 9.3(0.08) & 0.65 & 0.45 \nl
      & R & 21.35(0.01) & 46.8(0.23) & 19.64(0.01) & 9.1(0.07) & 0.72 & 0.47 \nl
      & I & \nodata     & 42.3(0.40) & \nodata     & 8.1(0.07) & 0.23 & 0.44 & a\nl
N5854 & B & \nodata     & 20.0(0.38) & \nodata     & 6.0(0.14) & 0.66 & 0.63 & a\nl
      & V & 20.15(0.22) & 19.5(0.35) & 19.07(0.23) & 5.7(0.12) & 0.55 & 0.61 & a\nl
      & R & 19.59(0.02) & 19.5(0.15) & 18.54(0.03) & 5.7(0.12) & 0.73 & 0.59 \nl
      & I & \nodata     & 19.6(0.17) & \nodata     & 5.6(0.11) & 1.05 & 0.63 & a \nl
\enddata
\tablenotetext{a}{Nonphotometric}
\end{deluxetable}

Recent literature
(e.g. de Jong 1996, Moriondo \etal ~1998) shows that for earlier type spirals,
such as those used in this sample, an $n=2$ bulge is the best fit. That finding
is supported likewise here. Furthermore, the
results indicate an increase in the B/D parameter with redder wavelengths,
compatible with the bulge having an overall redder color than the
disk and there being less obscuration from dust at longer wavelengths.  
A summary of the B/D parameters obtained 
for each filter and adopting the $n=2$ bulge is given in Table 4.

\subsubsection{Narrowband \Ha ~Images}

\Ha ~images were produced by observing the same region in two narrow (80\AA)
bandpasses, one centered on the appropriately redshifted \Ha ~emission line 
(the ON image) and a second just off the emission line (the OFF image).  
Both OFF and ON images were obtained over the nights of April 22--23, 1996
with the KPNO 0.9m telescope. The same T2KA chip, as described above, was used; 
however, only the central 1024x1024 pixel region was exposed. Usually, two ON exposures
of 30 minutes each and a single OFF one of the same duration were obtained;
Table 2 gives the details of the \Ha ~observations for each galaxy.
Images were trimmed, bias subtracted, flat fielded, and
cosmic ray edited in the same manner as the $BVR$ images described above.
The ON and OFF images were then convolved to the same resolution, and the two
ON images were combined. The OFF image
was scaled to the summed ON image and subtracted to produce
an \Ha ~line image.  The nights were both non--photometric and no
calibration is available.

\subsection{Optical Long Slit Spectra}

These data are described in more detail in Jore (1997).  Optical long slit
spectra were taken at the Palomar 5m Hale Telescope\footnote{Observations at
the Palomar Observatory were made as part of a continuing collaborative
agreement between the California Institute of Technology and Cornell
University.} using the red and blue cameras of the Double Spectrograph.  The
slit dimensions were 120\arcsec ~by 2\arcsec.  
A dichroic at 5500 \AA ~and gratings in both cameras
with dispersions of 1200 lines mm$^{-1}$ were used.  The red camera was
centered on the emission lines of \Ha ~and [N II] $\lambda$ 6548 \AA ~and
$\lambda$ 6584 \AA, with the [S II] $\lambda$ 6717 \AA ~and $\lambda$ 6731 \AA ~lines
also present.  The blue camera was centered on the Mg I b $\lambda$
5183.6, 5172.7 and 5167.3 \AA ~stellar absorption lines with the
[O III] $\lambda$ 5007 \AA ~emission line also visible.  Table 5 summarizes the Double
Spectrograph setup, while Table 6 presents the details of all spectral
observations used to extract rotation curves.  All exposures
over 40 minutes composed of co-added shorter exposures. 
It should be noted that some spectra were taken under conditions of poor 
transparency so that the effective integration is significantly less than 
the actual exposure time. Arc lamps and
radial velocity standard stars were observed for wavelength and velocity
calibration. In general, arc lamps were used for wavelength calibration
on the blue side, while night sky lines provided the necessary scaling
anchor on the red side.

\begin{deluxetable}{lllcccc}
\tablewidth{0pt}
\tablenum{5}
\tablecaption{Double Spectrograph Observing Setup \label{tabdbs}}
\tablehead{  
\colhead{Dates} & \colhead{Camera}  & \colhead{Chip} & \colhead{Spat. Res.} & \colhead{$\lambda$ Range} 
& \colhead{$\lambda$ Res.} & \colhead{Vel. Res.} \\[.2ex]
\colhead{} & \colhead{} & \colhead{} & \colhead{(\arcsec\ pix$^{-1}$)} & \colhead{(\AA)} &
\colhead{(\AA\ pix$^{-1}$)} & \colhead{(\kms)}}
\startdata
Mar. 94 - & Red & TI 167 & 0.58 & 6190-6850 & 0.82 &  38 \nl
\ \ Apr. 95  & Blue & TI 432 & 0.78  & 4950-5400 & 0.56  & 38 \nl
Mar. 96 & Red & 1396CR14-0 & 0.46 & 6150-6825  & 0.66 &  30 \nl
  & Blue & TI 432 & 0.78 & 4950-5400 & 0.56 & 38 \nl
\enddata
\end{deluxetable}

\begin{deluxetable}{llcccc}
\tablewidth{0pt}
\tablenum{6}
\tablecaption{Optical Long--Slit Spectral Observations}
\tablehead{  
\colhead{Object} & \colhead{Date}  & \colhead{Total Int.} & \colhead{PA} & \colhead{Axis} \\[.2ex]
\colhead{} & \colhead{} & \colhead{(seconds)} & \colhead{(\arcdeg ~E of N)} & \colhead{}}
\startdata
NGC 3626 & April 1995 & 3600 & 157 & major \nl
         &            & 1800 & 67  & minor \nl
NGC 3900 & March 1994 & 3600 & 2   & major \nl
         &            & 3600 & 92  & minor \nl
         & April 1995 & 3600 & 2   & major \nl
         &            & 1800 & 92  & minor \nl
NGC 4772 & April 1995 & 3600 & 147 & major \nl
         &            & 1800 & 57  & minor \nl
NGC 5854 & March 1994 & 3600 & 55  & major \nl
         &            & 3000 & 145 & minor \nl
         & April 1995 & 3600 & 55  & major \nl
         &            & 1800 & 145 & minor \nl
         & March 1996 & 6419 & 55  & major \nl
\enddata
\end{deluxetable}

Rotation curves were extracted for each of the ionized gas components using a 
Gaussian fitting routine. The stellar rotation curves were extracted using
several different methods employing both cross--correlation and
direct--fitting techniques. Further discussion of the reduction procedure is 
presented in Jore \etal ~(1996) and Jore ~(1997). The gas rotation curves displayed in
Section 3 are derived from \Ha ~and/or [NII], depending on the relative
strengths of the lines and the possible confusion with line absorption in
the case of \Ha. The stellar rotation curves are discussed here; the
stellar velocity dispersions will be discussed elsewhere (Jore \etal,
in preparation).

\subsection{HI Synthesis Imaging}

The \HI ~21--cm line observations were conducted with the Very Large Array
(VLA)\footnote{The Very Large Array is a facility of the National Radio
Astronomy Observatory.} in its
C configuration. As discussed in
Section 3.4, NGC~5854 is very gas--poor and was not mapped. 
A summary of the observational setup for the VLA observations of
the other three galaxies is given in Table 7.
Each galaxy was allocated 8 hours of array time in its C configuration,
with a nominal beam size of $\sim$12\arcsec ~and sensitivity to structures
smaller than 7\arcmin. Within each 8 hour allocation, a
flux calibrator was observed three times for $\sim$10 minutes, and a
phase calibrator was observed for 4 minutes between each 50 minute
observation on the galaxy.   All observations were
obtained using the 4ABCD spectral line mode with two overlapping
spectral bands of 32 channels per band, with an overlap
of at least 10 channels so that the total number of channels was 54,
each 20.6 \kms ~wide.
The total amount of on--source integration time and the 
total velocity coverage of the combined spectral bands are 
listed in Table 7.

\begin{deluxetable}{lccccl}
\small
\tablewidth{0pt}
\tablenum{7}
\tablecaption{VLA--C HI Synthesis Observing Parameters \label{tabvla}}
\tablehead{
\colhead{} & \colhead{weighting} & \colhead{N~3626} & \colhead{N~3900} & \colhead{N~4772} \\[.2ex]
\colhead{} & \colhead{scheme} &
\colhead{U~6343} & \colhead{U~6786} & \colhead{U~8021}}
\startdata
Date Obs.                &  & 2/23/96 & 2/24/96 & 11/10/94 & \nl
T$_{\rm int}$ (min)      &  &  373.5   & 371.5   & 403 & \nl
Velocity Coverage (\kms) &  & $ 930 - 2012$ & $1251 - 2336$ & $ 485 - 1564$ & \nl
Beam Size (\arcsec)      & uniform &  12.1 x 11.8 & 11.7 x 11.7 & 13.8 x 12.4 \nl
                         & natural &  16.8 x 16.2 & 22.5 x 20.0 & 31.7 x 23.1 \nl
Beam PA (\arcdeg)        & uniform& 113 & 165 & 143\nl
                         & natural & 101 &  79 & 137 \nl
RMS channel$^{-1}$ (mJy~beam$^{-1}$)  & uniform& 0.28 & 0.36 & 0.44\nl
                                      & natural & 0.23 & 0.31 & 0.33 \nl
minimum N$_{HI}$ (10$^{20}$ atoms~cm$^{-2}$)  & natural & 1.0 & 0.3 & 0.2\nl
\enddata
\end{deluxetable}

The UV data were flux and phase calibrated using the AIPS software
package after flagging bad visibilities.  The bandpass for each
antenna of the line data was calibrated for each IF separately.  The
two spectral bands were then combined using the task UVGLU.  Dirty
maps were made using MX and were inspected to determine the channels
free of line emission.  The continuum was subtracted in the UV plane
using UVLIN by interpolating the continuum from the emission-line free
channels to those with line emission.  Cleaned map cubes were created
from the UV data using a Gaussian beam
derived from the FWHM of the dirty beam.  For most of the galaxies,
two maps with different resolutions were made; one (using natural 
weighting) with lower spatial
resolution and higher signal-to-noise, and the other 
(using uniform weighting) with higher
resolution but noisier.  The beam size and position angle and
RMS noise per channel for each map are listed in Table 7.

Once the map cubes were made, further analysis of the data was done
using the GIPSY (van der Hulst \etal ~1992) data reduction package.  The
data cubes were first smoothed to twice the original beam size and
then clipped at the 1$\sigma$ level.  The regions containing line
emission in two or more consecutive channels were interactively
extracted.  The data in the original cube corresponding to the regions
extracted in the smoothed cube were conditionally transferred.  This
final cube contains only the line emission from the sources.  Zeroth
and first order moment maps (integrated intensity and velocity field
maps) were made using the GIPSY task MOMENTS.

Column density maps were obtained from the integrated intensity maps
using the relation \begin{equation}
N(H)=1.10297\times 10^{24} (1+v/c)^2 {1 \over a b}\int
S(v) dv\ \ [{\rm atoms\ cm}^{-2}]
\end{equation}
where $\int S(v)dv$ is the integrated flux (in Jy/beam) times the velocity
width (in \kms/channel), $a$ and $b$ are the FWHM of the major and minor axes of
the beam (in arcsec), and $v$ is the systemic velocity of the source (in \kms). The minimum
detectable column density quoted in Table 7 is the 2$\sigma$ value
in the natural weight column density map.

\subsection{Continuum emission}
Cleaned maps were made of the non--continuum subtracted maps.  The
continuum map for each source was determined by taking the mean of all
line--free channels.  The primary beam corrected flux density of nearby
sources was calculated using the GIPSY routine FLUX. The continuum
results for each galaxy are discussed separately; the derived continuum
flux density is included in Table 8.

\subsubsection{NGC~3626}

For NGC~3626, the average of 17 line--free channels was calculated, yielding a
continuum map. Continuum emission was detected at the center of
NGC~3626 with a flux density $S_{1413}=9.9$~mJy, slightly larger than
the flux density $S_{1415}=6$~mJy measured by van Driel \etal ~(1989).
There is another nearby source 
located 2\farcm9 west of the center (11\hrs17\mts13\fs5, 
+18\arcdeg37\arcmin38\arcsec ~(1950)) with a flux density
$S_{1413}=4.1$~mJy.  The companion galaxy, UGC 6341, has no continuum
emission associated with it, at a 3$\sigma$ point source sensitivity
of 0.3 mJy.

\subsubsection{NGC~3900}

For NGC~3900, the continuum map was constructed similarly.
Continuum emission was detected at the center of
NGC~3900 with a flux density $S_{1412}=1.2$~mJy in agreement with 
the upper limit of $S_{1415}<6$~mJy given  by van Driel
\etal ~(1989).  A nearby source is located 2\farcm8 north of
the center (11\hrs46\mts35\fs3, +27\arcdeg20\arcmin24\arcsec 
~(1950)) with a flux density $S_{1412}=1.4$~mJy which is
coincident with a small background galaxy seen in the
optical images.

\subsubsection{NGC~4772}

For NGC~4772, the average of 16 line-free channels was used to produce the
continuum map.  Continuum emission was detected at the center of NGC~4772
with a flux density $S_{1415}=3.4$~mJy.  Four other nearby
sources were found. Their positions and flux densities are as follows:
(a): 12\hrs50\mts59\fs9, +02\arcdeg24\arcmin43\arcsec ~(1950) with a
flux density $S_{1415}=5.0$~mJy; (b): 
12\hrs51\mts06\fs0, +02\arcdeg26\arcmin39\arcsec ~(1950) with a 
flux density $S_{1415}=4.0$~mJy;
(c): 12\hrs51\mts09\fs5, +02\arcdeg28\arcmin15\arcsec ~(1950)
with a flux density $S_{1415}=2.3$~mJy; and (d):
12\hrs50\mts58\fs3, +02\arcdeg29\arcmin35\arcsec ~(1950)
with a flux density $S_{1415}=3.7$~mJy.

\subsection{HI ~Channel Maps and Global Profiles}

The channels maps extracted from the natural weight data cubes
are shown in Figures 1 to 3 for the individual galaxies NGC~3626,
NGC~3900 and NGC~4772 respectively. The global
HI profiles extracted for the three target galaxies and for two others, 
UGC~6341 and NPM1G+18.0219, both found in the field of NGC~3626,
are shown in Figure 4. Table 8 presents the properties of these
five objects as derived from the VLA maps; NGC~4138 is also included
for comparative purposes. Entries are: 
(1) the continuum flux density
derived from the VLA map as discussed in Section 2.4, in mJy;
(2) the \HI ~deficiency, $<DEF>$, given by Magri (1994). $<DEF>$
is the logarithmic difference between the observed \HI ~mass and that expected
for an isolated galaxy of the same morphological class and linear {\it optical}
diameter (Haynes \& Giovanelli 1984); 
(3) the \HI ~line flux $\int S_{HI}~dV$ in Jy--\kms. For the VLA datasets, 
$\int S_{HI}~dV$ is derived from the natural weight data cube, after application
of a primary beam correction. 
(4) the \HI ~mass, derived from $\int S_{HI}~dV$, in \msun;
(5) the \HI ~mass to blue luminosity ratio, $M_{HI}/L_B$, in solar units;
(6) the \HI ~line heliocentric systemic velocity, $V_{21}$, in \kms;
(7) the width of the \HI ~line global profile at 20\% of the peak
intensity, $\Delta V_{21}$, in \kms;
(8) the ratio of the \HI ~to optical radius, $R_{HI}/R_{25}$, where
$R_{HI}$ is measured at a constant surface density of 
1 \msun~pc$^{-2}$;  
(9) the position of the dynamical center;
(10) the inclination derived from the tilted ring model, in degrees;
(11) the maximum rotational velocity derived from the best-fit rotation
curve, $V_{max}$, in \kms.
Further discussion is deferred
to the next section in which each galaxy is discussed individually.

\begin{deluxetable}{lccccccc}
\small
\tablewidth{0pt}
\tablenum{8}  
\tablecaption{\HI ~Properties \label{tabHIprop}}
\tablehead{
\colhead{} & \colhead{} &  \colhead{N~3626} & \colhead{NPM1G} & \colhead{N~3900} & \colhead{N~4138} & \colhead{N~4772} & \colhead{N~5854} \\[.2ex]
\colhead{} & \colhead{U~6341} & \colhead{U~6343} & \colhead{+18.0219} & \colhead{U~6786} & \colhead{U~7139} & \colhead{U~8021} & \colhead{U~9726}}
\startdata
S$_{cont}$ (mJy)          & $<$ 0.3  &  9.9$\pm$0.1 & $<$ 0.3 &  1.2$\pm$0.1 & 32\tablenotemark{b} & 3.4$\pm$0.2 & \nodata \nl
$<DEF>$\tablenotemark{a}  & \nodata  & +0.07 & \nodata & +0.02 & +0.08 &  +0.20 & 1.62 \nl
$\int S_{HI}~dV$ (Jy~\kms)    & 2.0$\pm$0.2  & ~8.3$\pm$0.3 & 1.7$\pm$0.2 & 16.7$\pm$0.7 & 20.6\tablenotemark{b} & 15.1$\pm$0.6 & 0.30\tablenotemark{c}\nl
$M_{HI}$ ($10^9$ $M_{\odot}$) & 0.27 &  1.1  & 0.23 & 2.9 & 1.2\tablenotemark{b} & 0.9 & 0.04\nl
$M_{HI}/L_B$  ($M_{\odot}$/$L_{\odot}$)  &  \nodata   & 0.06~  & \nodata & 0.2 & 0.2 & 0.1 & .01 \nl
$V_{21}$ (\kms)           &  1645    & 1484  & 1466 & 1801 &  888\tablenotemark{b} &  1040 & 1654\tablenotemark{c} \nl
$\Delta V$ (\kms)\tablenotemark{d} & 139  & 382 & ~96 & 450 & 340\tablenotemark{b}  &  463 & 128\tablenotemark{c} \nl 
$R_{HI}/R_{25}$           & \nodata  &  2.6  & \nodata &  1.8\tablenotemark{e} & 2.5\tablenotemark{b} & 1.9 & \nodata \nl
Dynamical center (1950)   & 111722.8 & 111726.1  & 111807.3 & 114633.9 & 120658.3\tablenotemark{b} & 125056.0 & \nodata \nl
                          & +183202  & +183748   &  +182938 & +271757  & +435749\tablenotemark{b} & +022621  & \nodata \nl
Inclination ($^\circ$)\tablenotemark{f} & \nodata  & 44--68 & \nodata  &  ~65 &  46--25\tablenotemark{b} & 65--58 & \nodata \nl
$V_{max}$ (\kms)                        & \nodata  &   195  & \nodata  &  220 &  208\tablenotemark{b} & 240 & \nodata \nl
\enddata
\tablenotetext{a}{HI deficiency measure from Magri (1994).}
\tablenotetext{b}{From Jore \etal ~(1996).}
\tablenotetext{c}{Based on reanalysis of digital archive copy of Magri's spectrum.}
\tablenotetext{d}{Width at 20\% of peak of global profile, uncorrected for smoothing or turbulence.}
\tablenotetext{e}{For main disk; including the southern extension, $R_{HI}/R_{25}$ $\sim$ 4.5}
\tablenotetext{f}{Where a range is given, a flat rotation curve has been assumed, and the range indicates
the variation from the inner to the outer disk. See Jore (1997).}
\end{deluxetable}

\section{Results for Individual Objects}

\subsection{NGC~3626}

NGC~3626 is found in an extended cloud of galaxies related to the NGC~3607 group,
LGG 237 (Garcia 1993); the central concentration around NGC~3607 is located about 
50\arcmin ~(350 kpc at the adopted distance of 24~Mpc) away. The only object known 
to be associated with the group that lies within 30\arcmin ~of NGC~3626 is 
UGC~6341, a Magellanic spiral at similar redshift, located at a separation of 
5\farcm9, 41 kpc. UGC~6341 was also detected in the VLA \HI ~synthesis map
and is discussed further below. A ten minute exposure with the Palomar 5m telescope
and the Double Spectrograph of the faint galaxy located
at R.A.(1950) = 11$^h$17$^m$30\fs1, Dec.(1950) = +18\arcdeg34\arcmin31\arcsec 
~failed to reveal any lines and its redshift is otherwise unknown.

Analyzing long--slit spectra of both the stars and ionized gas,
Ciri \etal ~(1995; see also Jore 1997) found that all of the gas in this galaxy is 
counterrotating relative to the stars. Recent CO observations by Garc\'{i}a--Burillo 
\etal ~(1998) reveal the presence
of a massive molecular nuclear disk associated with the ionized gas.
The \HI ~has been mapped previously by van Driel \etal ~(1989) with the
Westerbork Synthesis Radio Telescope (WSRT).

As evident in Figure 5, NGC~3626 is a ringed Sa galaxy with smooth, faint spiral 
structure and a prominent dustlane to the west of the nucleus. The three left panels
of Figure 5 show different representations of the $R$ band image, displayed with different
contrast levels and scaling to emphasize the low surface brightness outer disk (top), the
ring (center) and the dustlane (bottom). 
Unfortunately, the telescope suffered poor tracking when the $B$ and $V$ images were
obtained but, while the PSF is degraded, colors can still be derived. 
The $B-V$ color image shows no large scale structure in the disk aside
from the dustlane.  The reddening seen in both $B-V$ and $B-R$ at
+10\arcsec ~along the minor axis coincides with the dustlane.
The bulge is small, dominating the inner R $<$ 9\arcsec, with a $B-V$ color of 
$\sim$ 0.7; two larger blue ($B-V$ $\sim$ 0.4) components are found about 6\arcsec ~to 
the NW and SE of the bulge. As shown in the top right panel of Figure 5,
the $R$ band surface brightness profile is well fit by an $n = 2$ bulge, yielding
a modest B/D$_R \sim$ 0.37. Similar values are obtained for the other
bands as given in Table 4. Structure seen in the surface brightness profile coincides
with the crossing of the dustlane and ring.

Zooming on the central arcminute, the lower right panels in Figure 5
show the uniform weight \HI ~column density map superposed on the  
$R$ band image (center right) and on the \Ha ~image (bottom right). The grayscale  
in the $R$ band image is chosen to highlight the dustlane west of the nucleus.
As seen in Figure 5, \Ha ~is detected only in the central regions, peaking 
at the nucleus and also in a ring of radius $\sim$15\arcsec. The ring of star 
formation is coincident with a shoulder on the R-band profile between 12 and 20\arcsec. 
The \HI ~in the inner regions avoids the nucleus but follows closely the \Ha 
~within the limits of the VLA map resolution. Comparison of Figure 5 with Figure 1
of Garc\'{i}a--Burillo \etal ~(1998) reveals that the molecular gas peaks in
the same overall ring, but just where the \HI ~and \Ha ~are not seen, and vice versa.

Figure 6 shows representations of the \HI ~distribution and velocity field
extracted from the low resolution (natural weight) column density map. 
In the three panels, the optical
boundaries D$_{25}$ $\times$ d$_{25}$ are indicated by the ellipse,
oriented along the optical major axis at PA = 157\arcdeg.
The left panel shows the \HI ~contours and their corresponding grayscale.
The center panel shows the \HI ~column density contours
superposed on a grayscale representation of the $R$ band image with the
same scaling as in the center left panel of Figure 5.
The \HI ~is found in the central regions at R $<$ 60\arcsec ~and in a low column 
density ring of N$_H \sim$ 1-3 $\times 10^{20}$ atoms cm$^{-2}$ peaking
at R $\sim$ 190\arcsec. This outer ring is
both exterior to the optical radius, significantly more elongated,
and oriented with a different position angle, nearly north--south.
The ratio of \HI ~to optical size is large: $R_{HI}/R_{25}$ = 2.6.
The rightmost panel
shows the \HI ~velocity field derived from the natural weight data cube.
The velocity field implies fairly regular rotation in the 
outer ring, with the north side of the galaxy 
is approaching and the south, receding. Many of the contours are
jagged due to the patchy distribution of the gas.  It is also apparent
that the inner \HI ~disk has a significantly different kinematic position
angle than that of the outer disk or as measured from the overall gas 
distribution.

Two nearby galaxies were detected in the field of view of the \HI~
map:  UGC~6341, which is located 5\farcm9 south
of NGC~3626, and a faint object identified as NPM1G +18.0219
(Klemola \etal ~1987) 12\farcm9 ~(90~kpc) to the SE. In fact, the 
highest \HI ~column density in the field is found coincident with the 
faint companion UGC~6341, an inclined low surface brightness Magellanic spiral. The \Ha
~image did not extend out this far but in the broadband images, two central knots are
visible which makes fitting central ellipses impossible. There is no bulge evident.
One of the knots appears to be the center of light (nucleus) and the other lies
about 10\arcsec ~to the NW. We derive m$_R$ = 15.04 $\pm$ 0.04, with $\epsilon$ 
= 0.72 and PA = 153\arcdeg. The global HI
properties of these objects are summarized in Table 8 and the
global \HI ~profiles of all three objects are shown in Figure 4.

In the central regions, the uniform weight \HI ~map is deep enough to show the 
column density contours, but the S/N is too low to trace the velocity field.
Within the resolution, the dynamical center coincides with the nucleus,
but the inner velocity field is clearly distorted.
To the north, the velocity contours close at R $\sim$12--25\arcsec ~but the velocity
begins to rises again outside 45\arcsec. 
This rise and subsequent fall appears to coincide with that seen in the
optical emission line gas rotation curve, shown in Figure 7. To
the south, even in the center, the velocity contours change continuously.
Analysis of the velocity field was accomplished by following the common
practice of fitting a tilted ring model, assuming circular rotation.  
In the center, the position angle of the \HI ~is closely aligned with
the optical axis, starting at 160\arcdeg ~but swiftly increases
to 195\arcdeg ~at a radius of 65\arcsec.  The position angle then
drops to 175\arcdeg ~by a radius of 140\arcsec ~and changes little
to the edge of the \HI ~disk.  Figure 7 shows position--velocity diagrams 
extracted from the natural weight data cube at position angles of  
157\arcdeg ~(upper left), along the optical major axis, and 
178\arcdeg ~(upper right), along the kinematic major axis of the outer 
\HI ~ring. In the left panel, the optical stellar and emission line velocities
derived from the long-slit spectra 
are superposed. It is clear that the \Ha ~and \HI ~gas show similar rotation curves 
in the inner part, but that both counterrotate with respect to the stellar
component. Moreover, it is also clear that the inner rotation curves of
the ionized and neutral gas are both distinct and shifted in orientation from that
traced by the outer \HI ~ring. The rotation curve of the \HI ~in the outer
ring along PA = 178\arcdeg ~continues to rise to its outermost
point; the \HI ~velocity gradient of the outer ring is minimized along
PA =90\arcdeg. A large velocity gradient is evident within the inner
$\sim$20\arcsec ~in all cuts through the \HI ~cube, due, at least in part,
to beam smearing. 

The \HI ~velocity contours suggest
that the inclination of the \HI ~disk varies considerably.  
The best fit inclination given by the tilted ring model
starts at 44\arcdeg ~and then drops to 33\arcdeg ~out to a radius of
80\arcsec. However, this drop is likely an artifact of the disconnect
between the inner \HI ~ring and the outer \HI ~ring; there is no \HI ~in this
region to constrain the velocity field.  Towards the outside, the inclination
appears to increase continually with radius until it reaches
a maximum of 68\arcdeg ~at the edge of the \HI ~disk. That the outer ring
is more highly inclined that the stellar disk is also suggested by
the differences in ellipticity seen in the isophotes of optical light and
\HI ~column density. With this variation in inclination, the rotational 
velocity remains relatively constant at 195 \kms ~throughout the \HI ~distribution.

Van Driel \etal ~(1989) obtained an \HI ~synthesis map of this object
with the WSRT. The signigicantly lower resolution (13\arcsec ~$\times$ 
41\arcsec) and sensitivity of that map did not allow discrimination of the
structural details but there is rough correspondence to the features
reported here.  In fact, despite the poor resolution of their derived
velocity field (40\arcsec ~$\times$ 41\arcsec), van Driel \etal 
~noted the marked shift of position angle
between the optical major axis and the morphological and kinematic
position angles of the outer \HI ~ring.

Ciri \etal ~(1995) obtained both absorption and emission line spectra at four
position angles: 25, 68, 113 and 158\arcdeg. They found that the stellar component
entirely counter--rotates with respect to the stars. Along the PA = 158\arcdeg
~axis, the stellar lines were traced to $\pm$30\arcsec, while the ionized lines
extended to $\pm$20\arcsec. The Palomar spectra were obtained
along PA = 157\arcdeg ~(major axis) and PA = 67\arcdeg ~(minor axis);
as shown in the lower panels in Figure 7, the extracted rotation curves for
both components are qualitatively similar to those shown in Figure 1 of
Ciri \etal ~but extend further in the spatial dimension. 
The stellar curves rise with some
deviation at about 8\arcsec ~to flatten at R $\sim$20\arcsec ~and
beyond. The stellar curves can be traced in these spectra to 45\arcsec ~on either
side of the major axis; this radius is only slightly more than 0.5~R$_{25}$. 

Emission line rotation curves can be traced to $\pm$ 20\arcsec ~in
[OIII] and [SII] and to $\pm$ 30\arcsec ~in \Ha ~and  [NII].
Although in the very center the \Ha ~is likely affected by absorption,
there appears to be a hint that the \Ha ~emission retains its high velocity almost
to the center, showing velocities of $\sim$150--180 \kms~ within the inner
5\arcsec. In fact, Ho \etal ~(1997a) classify NGC~3626 as a possible LINER, based on
line ratios; the FWHM of the [NII] within their window along PA = 170\arcdeg
~(Ho \etal ~1995) 
is large, 413 \kms. The [NII], [SII] and [OIII] lines
show steeply rising velocities along the major axis, but also evidence
of line broadening or a secondary component is seen at symmetric
points within R $\le$ 6\arcsec. In particular, the [NII] lines appear to double
within the innermost region and have been fit by two separate Gaussian components,
as indicated in Figure 7. When fit separately, the second component appears to
match the velocity of the stellar component.
In addition, both the \Ha ~and [NII] curves peak at
symmetric radii at R $\sim$ 10--15\arcsec~ but then show a clear decline 
outside R $\sim$ 20\arcsec. This decline is suggested but
not evident in the lower sensitivity data of Ciri \etal ~(1995). 
As discussed above, these features are also reflected
in the \HI ~velocity field. 
The small deviation from constant velocity seen in the stellar values along the 
minor axis is mostly likely due to a slight offset from the true dynamical center 
(i.e. pointing) or minor axis position angle. Along the minor axis, 
both the [NII] and [SII] spectra exhibit double--peaking within the inner 
R $\sim$ 6\arcsec. Within the resolution limits, it is
uncertain whether the line broadening is merely a LINER phenomenon, excess streaming, or
a secondary gas component co--rotating with the stars. Garc\'{i}a--Burillo \etal 
~(1993) likewise find suggestive evidence for a secondary gas component associated with
the dominant stellar disk in the central regions.

In summary, the kinematic evidence for NGC~3626 suggests a large scale
decoupling of the stars and gas. The \HI ~distribution is found in two
kinematically distinct and spatially separate rings that show a large difference 
in position angle and inclination. Both the morphological and kinematic position
angles of the outer \HI ~ring are shifted by $>$ 20\arcdeg ~with respect to
starlight and the inner \HI ~gas. Evidence of a gas component co--rotating 
with the stars is suggested, as hinted also in the CO map
of Garc\'{i}a--Burillo \etal ~(1998). 

\subsection{NGC~3900}

NGC~3900 is a ringed Sa galaxy with no nearby companions. The nearest known
object of similar redshift,  UGC~6791, lies to the SE at a projected separation of 
17.1\arcmin, 135 kpc at the adopted distance of 27 Mpc. NGC~3912, also
considered part of a group LGG~252 by Garcia (1993) lies 34.8\arcmin
~(275 kpc) to the SW.  There are several small galaxies nearby;
two of them are included in the catalog of dwarfs found by Binggeli \etal ~(1990).
None of these possible companions has a known redshift. For one, \#44 of 
Binggeli \etal, we failed to detect any lines in a 600 sec exposure
obtained with the Palomar 5m plus Double Spectrograph using the setup described
in Section 2.2.

As evident in the $B$ band images shown in Figure 8, NGC~3900 shows a small circular
bulge with a prominent blue inner ring at a radius of 34\arcsec. In the 
exterior, the light from the disk abruptly decreases at roughly half the optical 
radius. Flocculent spiral structure is clearly evident in a diffuse, faint disk 
that, as seen in the upper panel of Figure 8, appears to be more elongated 
than the higher surface brightness portions, as might be expected for 
a tightly wound spiral seen at moderate inclination. Only a
hint of \Ha ~is evident in the narrow-band images with a single concentration
near the center and a V--shaped structure in the north corresponding
to blue enhancements ($B-V$ $\sim$ 0.6) in the northern section of the ring
evident also in the color maps. A dustlane is seen on the southern side 
looping to the east and wrapping nearly halfway around
the galaxy.  The dustlane is easily visible on the northern and southern ends
in the $B$ band image, just interior to the inner ring and more easily distinguished
in the east where it overlaps the inner ring in the color map images.  

From examination of a HST WFPC2 F606W image,
Carollo \etal ~(1997) classified N~3900 among the early
spirals with  a ``classical spheroidal bulge''; the bulge component 
identified in the HST has a smaller effective radius than the limited
resolution ground based images could distinguish from a central component.
In fact, as evident in Figure 8, a $n$ = 2 bulge fits quite well 
and gives a modest B/D$_R$=0.30. Structure in the surface brightness profile
is evident in all the broadband images where crossings of the ring and
dustlane occur, and where the outer faint disk contributes.
Use of a higher order bulge in our $R$ band
image yields a poorer fit with a higher B/D$_R$=0.71 in which the bulge
clearly overshoots the observed central light profile; the $n=2$ bulge also
provides a better fit to the outer (R$>$ 80\arcsec) disk.

NGC~3900 was mapped previously with the WSRT by van Driel \etal
~(1989).  In their map, most of the \HI ~is found in a ring of radius
1\arcmin, with some suggestion of the western extension. This ring
is clearly evident in Figure 9 which shows different representations
of the \HI ~map and velocity field extracted from the natural
weight VLA data cube. The \HI ~indeed peaks in a ring at R $\sim$ 50\arcsec,
just outside the prominent blue ring, and in the region dominated
by flocculent spiral structure.
The lower sensitivity of the WSRT map
completely missed the extended diffuse structure to the north and
south, evident here in Figure 9. In addition to the higher density
\HI ~which is roughly coincident with the optical galaxy, diffuse \HI~
can also be traced in patches to the north and south. To the south, an extension
is visible, clumping about 5\arcmin ~(40 kpc) from the optical center. 
Furthermore, the \HI ~column density
distribution associated with the main body is clearly asymmetric, extending
nearly twice as far to the west as to the east.

The velocity contours shown in the right panel of
Figure 9 and the top left panel of Figure 10 
clearly suggest that the southern gas cloud is simply
an extension of the main \HI ~distribution, the column density dropping
below the sensitivity limit of the VLA map in between. In the inner
parts, the contours in the \HI ~disk are very symmetric about the major axis
and show regular rotation. The main \HI ~disk extends only
2\farcm9 (23 kpc) from the center; along the major axis,
$R_{HI}/R_{25}=1.8$.  However, comparing the total extent of \HI ~gas
on the southern side of the galaxy to the stellar disk 
gives $R_{HI}/R_{25}=4.5$; the \HI ~in the south
can be traced to R $\sim$ 50 kpc.  Along the minor axis, to the west,
the \HI ~extends about twice as far as to the east
and deviates in velocity as would be expected if the \HI ~disk were 
strongly warped in the outer parts.

The velocity field was modeled using tilted rings and including the southern 
low column density gas under the assumption that it lies in the same
plane, at a constant inclination of 65\arcdeg.
In the best--fit model, the position angle
varies from 178\arcdeg ~to 186\arcdeg ~in the main \HI ~disk and
then settles back to 176\arcdeg ~at a radius of 200\arcsec.  With these 
parameters, the
rotational velocity rises to $\sim$200 \kms ~in the inner 60\arcsec
~and is then relatively constant except between 240\arcsec ~and
300\arcsec ~where the parameters are poorly constrained because there
is little gas in that region. In these models, the western extension is not
well fit because of the asymmetry in the distribution along the minor axis.

Figure 10 shows the position--velocity diagrams for the \HI ~distribution
along both the optical major (PA = 2\arcdeg) and minor (PA = 92\arcdeg) axes;
the velocities extracted from the long--slit spectra for stellar and gaseous
components are superposed. The optical and \HI ~rotation curves are similar,
but the optical ones do not conclusively show the flattening seen in the
\HI ~data. There is also a suggestion of a spread in the velocity along
PA =2\arcdeg ~to the north at R $\sim$ 1\farcm5, showing a blueshifted cloud 
with $\Delta$V $\sim$100 \kms. 
In the upper right panel of Figure 10, the minor axis 
position--velocity diagram shows clear evidence of the \HI ~ring
and central hole. It also reveals a small change in velocity along the western
\HI ~extension, likely due to the proposed warp.   

As illustrated in the lower panels of Figure 10, the
optical spectra yield both gas emission and stellar absorption lines
extending along the major axis to R $\sim \pm$ 55\arcsec. The stellar
\MgIb ~lines are relatively weak outside the bulge dominated region,
at radii $>$ 10\arcsec ~to the north and $>$ 15\arcsec ~to the south.
The stellar component to the N of the nucleus shows an asymmetric line
of sight velocity distribution
that could be either two separate components but is more likely simply
due to true asymmetric deviation from a Gaussian distribution;
this detail will be discussed elsewhere (Jore \etal ~in preparation). 
The central \Ha ~is strongly affected by absorption within
the inner 20\arcsec~ to the north or 15\arcsec~ to the south,
but matches nicely the [NII] further out. In the inner rotation curves
derived from the [NII], [OIII] and [SII] lines, a significant anomaly,
indicating blueshifts of $\sim$160 \kms, is seen,
peaking $\sim$5\arcsec~ to the north of center along the major axis.
Beyond this radius, the velocities fall and resume the expected rise
seen to the south at about 10\arcsec. There is a suggestion that the
\NII~ line is double peaked in the center.  Ho \etal ~(1997a) classify 
NGC~3900 as a possible LINER based
on line ratios extracted from a spectrum obtained along a PA of
178\arcdeg ~(Ho \etal ~1995). However, as can be seen in the lower
panel of Figure 10, the [NII] lines exhibit a systematic sinusoidal
wobble within $\pm$7\arcsec~ from the center along the minor axis.
While this behavior might be expected from a small offset in position
or position angle from the true kinematic center, the stellar
velocities, extracted from the same spectra, remain flat. 
The color maps show a coincident feature  --- slightly bluer 
on the east and redder on the west of the nucleus at comparable radii. 

While the main disk of NGC~3900 shows motions dominated by
circular rotation in a quiescent pattern, anomalous emission line 
velocities imply non-circular motions in the central region. Interior to
the optical edge, the \HI ~is found in a ring or tightly
wound spiral pattern with symmetric peaks to the north and south
along the major axis in the region of flocculent optical
spiral structure. At low column densities, 
N$_H \le$ 2 $\times 10^{20}$ atoms cm$^{-2}$, the \HI ~distribution
is huge. The outer \HI ~can be traced to a radius more than 4
times the optical size in a diffuse and patchy distribution
that appears kinematically connected to the main galaxy.
Along the minor axis, the \HI ~distribution is both warped
and strongly asymmetric.  Models of the evolutionary history
of this galaxy must account for the kinematic quiescence of
the patchy and asymmetric \HI ~far beyond the optical edge.

\subsection{NGC~4772}

NGC~4772 is an outlying member of the Virgo cluster, probably associated
with the subclump around NGC~4472. The adopted distance assumes this
association and the derived subgroup distance of 16 Mpc from distance
moduli given in Ferrarese \etal ~(2000). Its nearest companion,
CGCG~015-036, a compact high surface brightness galaxy with a systemic velocity 
of \vsun ~= 844 \kms, is found 
~at 18\farcm5, 85 kpc at the adopted distance of 16 Mpc to the 
E--SE. The pair NGC~4809/4810 at similar velocity lies 35\arcmin
~(160 kpc) to the NE.

Dubbed the ``eye galaxy,'' a descriptive term adequately
describing its appearance as shown in Figure 11, NGC~4772's center is dominated
by a slightly flattened bulge surrounded further out by a segmented
ring.  The western and eastern sides of the ring form continuous structures;
however, the northern and southern ends are broken and patchy. 
The inner galaxy shows a clear dustlane surrounding the bright bulge
and oriented parallel to the major axis. The bright, red circular bulge, of
$B-V$ $\sim$ 1.0, is clearly separable 
within the inner 5\arcsec, cutoff on the north and east by the dustlane.
Structure (filaments or braids) are seen within the dustlane, which must
trace spiral structure, not simply a ring. The bulge--disk decomposition
yields a moderate B/D$_R$ = 0.44. 

The \Ha ~shows a central peak, coincident with the nucleus,
surrounded by an elongated (b/a $\sim$0.3) \Ha ~ring which itself 
is crisscrossed but entwined in the dustlane and which coincides
with the \HI ~column density peaks. 
As evident in the upper panel in Figure 11, there are two pronounced
drops in surface brightness, one at about 65\arcsec ~and a
second at about 150\arcsec. This latter coincides with the outer edge 
of the inner \HI ~ring, just about where there are suggestions of the
start of spiral arms. At that radius, the isophotes, $\mu_B$ = 25.9, 
$\mu_R$ = 24.0 \magarc, become rounder, changing from  
$\epsilon \sim$ 0.47 to $\epsilon \sim$ 0.29, similar to what is seen
in the \HI ~distribution. At the faintest isophotes, this rounder low
surface brightness distribution can be traced to more than 200\arcsec,
twice R$_{25}$, and roughly to the outermost \HI ~contours. The outer
low surface brightness disk is blue, $B-V$ $\sim$ 0.67 versus 
$B-V$ $\sim$ 0.75 -- 0.85 further in, but contributes
only a small amount of light (0.19 mag in B) to the total luminosity.
The faint optical light is visible on the DSS image but makes 
the galaxy rounder than published axial ratios suggest.  
It is reminiscent
of a similar feature detected by Buta \etal ~(1995) in deep
photometry of the counter-rotating Sab galaxy NGC~7217. 

As evident in Figure 12, the \HI ~gas is distributed in two distinct rings
surrounding a central hole.
Coincident with the \Ha ~ring evident in Figure 11, the inner high column
density \HI ~ring peaks at R $\sim$ 60\arcsec,
with a position angle of 144\arcdeg, similar to that of the bright
optical disk and with comparable eccentricity. Surrounding the central
hole, the full ring is characterized by moderate column densities,
N$_{H} >$ 7 $\times 10^{20}$ atoms cm$^{-2}$. Clearly separated from this
inner ring, the rounder outer \HI ~ring
can be traced to $\sim$ 200\arcsec ~in radius oriented along a position 
angle of 150\arcdeg ~and, as seen in the lower
right panel of Figure 12, extends over an area similar to, but of different
eccentricity, with the faint blue stellar light.  The outer ring is patchy,
but, in the north, can be traced over 90$^\circ$ in azimuth at N$_H \sim$
2 $\times 10^{20}$ atoms cm$^{-2}$, with localized patches of
3 $\times 10^{20}$ atoms cm$^{-2}$. The velocity field shows
a similar discontinuity.
Near the edge of the inner ring and in the outer ring, 
along the major axis, the isovelocity contours bend, suggesting 
a warp in the outer regions.  Figure 13 shows the
position-velocity diagrams along the optical major (PA=147\arcdeg) and
minor (PA=57\arcdeg) axes; the velocities derived from the optical
spectra in the same regions are shown. The
velocity rises linearly in the inner regions and then flattens out at
a radius of 90\arcsec. Within the outer ring, the \HI ~velocities
show a significant turn-over at large radii in the SE.  
Along the minor axis, the contours appear
regular and symmetric surrounding the central hole, 
but the gradient of the outer \HI ~velocities
reinforces the notion that the outer gas disk is warped. 

As in the cases of NGC~3626 and NGC~3900, the
fitting of tilted ring models is complicated
by the discontinuity in the \HI ~distribution between the
inner and outer rings. The inner region
shows a rotational velocity smoothly rising to 240 \kms ~at 
a radius of 100\arcsec. If a flat rotation curve, 
set by the fit to the inner disk, is adopted,
the position angle starts at 143\arcdeg ~and rises steadily to 160\arcdeg
~near a radius of 140\arcsec ~remaining roughly constant
at $\sim$158\arcdeg ~further out.  However, just at the edge
of the inner \HI ~ring, the inclination becomes unconstrained and
discontinuous. A second fit was performed
to smooth over the discontinuity in inclination by fixing the
inclination for the transition rings to allow a smooth variation
in the inclination. In that case, the inner rotational
velocity rises much in the same way as in the previous model, but
in the outer ring, it rises by 60 \kms. Yo maintain a flat rotation
curve, the inclination must drop outside R$>$ 110\arcsec ~from
about 65\arcdeg ~to about 58\arcdeg, as suggested also by the
rounding of the aspect ratio of the outer column density contours.
In terms of residuals,
both of these models fit the observed velocity field equally well.
Further details of the resultant mass modelling will be discussed
elsewhere.

Figure 13 also shows the stellar and ionized gas velocities extracted
from the long slit spectra. The stellar lines can be traced from about
60\arcsec ~on the SE to about half as far on the NW. The ionized gas
distribution is extremely patchy. [NII], [OIII] and [SII] emission
is detected within the central R $\le \pm$6\arcsec, and on the
southeast side along the major axis in clumps at R $\sim$ 25\arcsec ~and
50\arcsec. No ionized emission is detected to the NW.
The \Ha ~line is depressed in
the center due to absorption as far out as 15\arcsec ~but the outer
\Ha ~emission coincides nicely with that of the other ionized species.  
As also evident in the top panels of Figure 13,
the optical rotation curves do not extend suffiently far to sample
the rotation curve peak.
 
Ho \etal ~(1997ab) classify NGC~4772 as a Seyfert Sy1.9, based
on line ratios and the appearance of broadened \Ha ~seen in their
spectrum obtained along PA = 5\arcdeg. They also remark on the flat 
or double--peaked nature of the narrow lines. The \Ha ~line broadening 
is also evident in our spectrum but the other species, particularly [OIII],
show a systematic rise to $\sim$ 80 \kms ~at R$\sim \pm$ 6\arcsec,
more steeply and of opposite sign than the stellar rise in the same region.
It is important to note that, unlike Ho \etal, who chose to orient
the slit close to the parallactic angle, our setup placed the slit
along the optical axis at P.A. = 147\arcdeg. Furthermore, as evident in 
Figure 13, the minor axis [NII], [OIII] and [SII] velocities rise
rapidly in the inner 5\arcsec ~up to a velocity of 100 \kms ~and then
decrease out to 15\arcsec ~in radius. In this case, the [OIII] provides
the best tracer, showing a velocity peak along the minor axis at 4-- 5\arcsec
~followed by a decline back to the systemic by the time it disappears
at about 15\arcsec. At the same time,  the minor axis stellar velocities are
very flat (verifying that the slit was correctly aligned along the minor axis of the
{\it stellar} component). The kinematic signature suggests the presence of a 
misaligned disk or bar.
The only unique feature seen in the color maps is a flattening of the $B-R$
in the central region along the minor axis and a hint of smaller plateaus
in all the color maps along the major axis. These two color features occur on
comparable scales as the central rotation curve peculiarities.  

NGC~4772 may represent the end--state of a prograde merger in which the
transfer of angular momentum leads to an outward spread of the disk
(Quinn \etal ~1993). The steep rise of the ionized gas velocities along the 
minor axis and the apparent decoupling on gas emission along the major axis suggests
the presence of a misaligned embedded disk or bar. Reminiscent of
NGC~3626, the \HI ~is contained in two separate but concentric rings of
differing position angle and eccenticity. The inner ring coincides with the
sites of current star formation in the optically bright inner disk,
while the other rounder one, wholly outside R$_{25}$ but coincident with a
faint blue stellar component.

\subsection{NGC~5854}

NGC~5854 is a gas--poor Sa with no evidence of current 
star formation. Relatively isolated, it is likely an outlying
member of the NGC~5846 group
(Haynes \& Giovanelli 1991; Garcia 1993;  Zabludoff \& Mulchaey 1998).
Its nearest neighbor is CGCG 021--009 at 33\farcm1, 230 kpc
at the assumed distance of 24 Mpc. 
An illustrative summary of results is presented in Figure 14.
 
The $B$ band image shows an inclined galaxy with a
large elongated bulge and little apparent spiral structure. Some hint
of arms is seen in the two small stumps on the NE and SW edges of the bulge.
Unfortunately, the $B$ band image was taken under non--photometric
conditions so that the color information is limited to $V-R$. In the $V-R$ 
image, little structure is seen, although the galaxy
is clearly redder than the others. It was previously imaged by Balcells 
\& Peletier (1994) who find B/D$_R$ = 0.46 using a 1-D decomposition method 
and find similar flatness of the color profile.  As discussed below,
no \Ha ~is seen in the long slit spectra, but \Ha ~absorption is seen across 
the disk and can be traced almost as far as the \MgIb ~lines.
The galaxy is thus not currently forming stars, but its significant
population of A stars suggests a post--starburst state.
Of the galaxies discussed here, it has the highest B/D$_R$ $\sim$ 0.6
for $n = 2$, but it is clear that the $n = 4$ fit, illustrated
in Figure 14, gives a more appropriate  B/D$_R$ $\sim$ 0.8.
No \Ha ~image was obtained.

As reported by Magri (1994), the \HI ~mass of NGC~5854 is very low 
($\sim 3 \times 10^7$ \msun) 
and hence it was not feasible to map its \HI ~with the VLA.
The optical spectra are dominated by stellar absorption lines which can be
traced, by both \MgIb ~and \Ha, out to $\pm$45\arcsec, further than, but 
qualitatively similar to
the rotation curve displayed in Simien \& Prugniel (1997). Those authors note
V$_{max}= 128 \pm 8$ \kms ~for the stellar component in their spectrum
traced to $\sim$ 27\farcs5. We find a comparable value at that radius,
but a larger V$_{max}= 167 \pm 10$ \kms ~at R $\sim$40\arcsec.
The \MgIb ~rotation velocities
rise more shallowly than the \Ha ~lines do, possibly because of extinction.
The [NII] and [OIII] emission reveals a distinct counter--rotating gas component traceable 
from the center out to 6\arcsec~ on the NE side and 8\arcsec~ on the SW side. 
The maximum rotational velocity of this [NII] feature is 64 \kms ~at $\sim$4\arcsec~
NE of the optical center of light, the [OIII] appears in a ring coextensive with
the [NII] gas but the velocities are higher, reaching $\sim$ 100 \kms.
The stellar velocity profiles along the minor axis show a slight trend that
most likely arises because of a small pointing offset. The weak [NII] shows some
dispersion but is essentially flat, within the errors; no other ionized emission
is seen. 

Magri (1994) detected a weak and narrow \HI ~signal from NGC~5854
at Arecibo after more than 2 hours of on--source integration. 
A zoomed reproduction of Magri's spectrum extracted from our digital 
archive is shown in Figure 14. The systemic velocity of the \HI~
is 1654 \kms ~which matches the value of 1663 $\pm$10 \kms ~given by 
Simien \& Prugniel (1997). We have remeasured
the \HI ~line width in the digital spectrum and corrected it for smoothing, etc.
following Haynes \etal ~(1999b) to obtain a value, uncorrected for inclination and
turbulence, of W$_{21} \sim$95 \kms. This value is too small compared
to the full rotation width of the galaxy, but is comparable to the velocity 
spread seen in the [NII] in the inner disk.
A simple interpretation of the narrowness of the \HI ~profile is that
that the \HI ~and [NII] both arise in a small central disk 
of R $<$ 8\arcsec ~that counter--rotates relative to the stars.
At the same time, the primary (stellar) disk is seen in a post-starburst phase,
showing strong Balmer absorption throughout. 
The current gas--poor state of NGC~5854 may thus be result of the accretion
of a relatively dense satellite that has heated the disk and smoothed
the spiral structure, leaving only the counter-rotating gas core as
evidence of its occurrence. 

\subsection{Another Look at NGC~4138}

As mentioned previously, the first object mapped in \HI ~with the VLA as part
of this study was NGC~4138. 
A discussion of the optical and \HI ~spectral observations is
presented in Jore \etal ~(1996); the gas counterrotates with respect to the
primary stellar component, but a secondary stellar population, kinematically
coupled to the gas, is visible.
The amplitude of the crosscorrelation function of the stellar lines coincides
spatially with an \Ha ~ring, previously noticed by Pogge \& Eskridge ~(1987),
at a radius of $\sim$ 20\arcsec. The two stellar
components are characterized by distinct velocity dispersions, implying separate 
origins. As part of the current study, we added $B$, $V$, $R$ and \Ha ~images to the
dataset for this morphologically normal but kinematically unusual object.

As evident in the top left panel of Figure 15, NGC~4138 is a small bulged, smooth 
armed Sa galaxy with a prominent dustlane in the SE quadrant. A foreground star 
is located at a radial distance of 4\arcsec~ from the center along the major axis. 
As seen in the middle panels of Figure 15, the inner \HI ~forms a partial ring, 
coincident with the \Ha ~ring, on the north and east, surrounding a central hole.
In $B-V$, the galaxy shows a distinctly red bulge ($B-V$ $\sim$ 1.1) and a
blue ($B-V$ $\sim$ 0.6) ring located in the approximate position of the \Ha
~ring.  The blue ring is particularly strong on the northern and southern portions
of the galaxy with breaks in the ring on the east and west.  The eastern break
is due, at least in part, to the dustlane. The $V-R$ image shows similar structure.
As shown in the upper right panel of Figure 15, the bulge--disk decomposition 
gives a small value: B/D$_R$ = 0.17. 

There are several outstanding features in the color profiles which correlate well
with features in its rotation curve as presented in Jore \etal ~(1996). At
$\pm$13\arcsec, where the secondary component of stars
begins, until $\pm$35\arcsec ~when the component is no longer detectable in
the long slit spectrum, there is an obvious reddening in the major axis profiles.
There is an abrupt transition moving from the center into the ring on the
southeast side and an abrupt transition out of the ring on the northwest side,
There are no apparent variations in the color profiles along the minor axis
aside from a peak redder color in the center.

The relationship of the \HI ~gas reported previously by Jore \etal
~to the ring and the dustlane is demonstrated 
here in more detail in Figure 15. Indeed, the relationship of
the dustlane and \Ha ~ring to the \HI ~distribution seen in NGC~4138 is 
very similar to that seen in NGC~3626 (Figure 15 versus Figure 5), the other 
case of large--scale counter--rotation. Likewise, the \HI ~extent is
very large, $R_{HI}/R_{25}$ = 2.5 and, as discussed in Jore \etal,
the velocity contours along the minor axis suggest a strong warp
beyond R$_{25}$. In contrast to the \HI ~distributions
seen in NGC~3626 and NGC~4772, however, the outer HI does not form a
distinct and coherent ring, but rather is extremely patchy beyond the
optical edge.

NGC~4138 remains the sole example in our Sa sample of a galaxy 
with an extended counter--rotating stellar component. Nonetheless, it bears 
similarities to the other objects in terms of its morphological ``boring--ness'',
relative isolation, large, low surface density \HI ~extent, and
overall regularity of its velocity field.

\section{Discussion}

In this paper, we have presented a combined imaging and spectroscopic dataset
for four Sa galaxies; images are also included for a fifth, NGC~4138,
discussed previously by Jore \etal ~(1996). All of the galaxies show evidence
of departure from kinematic normalcy ranging from large--scale counter--rotation
to decoupled central gas and/or stellar components. Despite the
difference of detail, several unifying themes seem critical: (1) All 
of the objects are relatively isolated, morphologically normal,
unbarred Sa galaxies. (2) Rings are prominent features both in the optical 
broadband light and in the distributions of ionized and neutral gas. The 
apparent rings may also result from tightly wound spiral arms viewed 
at moderate inclination. 
(3) The kinematically decoupled gas components appear to be associated 
with sites of current or at least recent star formation. 
In all cases where \Ha ~emission is detectable in the
region of interest, it coincides well with the location of kinematically
decoupled components.  When there is no \Ha ~emission in these regions,
the decoupled components often show some form of a color change, often
towards the red. (4) The gas--rich galaxies (all except NGC~5854) 
contain moderate \HI ~masses, but,
because the gas is spread over a large area R$_{HI}$/R$_{25} \ge$ 2,
the globally averaged \HI ~surface densities
$<\sigma_{HI}> = M_{HI}/\pi R_{HI}^2$ are very low, typically of order 0.5--1.0 
M$_\odot$~pc$^{-2}$. In NGC~3626
and NGC~4772, the \HI ~is found in two concentric but distinct rings.
In NGC~3900, the gas exterior to the optical edge is patchy, but can
be traced 50~kpc to the south, giving R$_H$/R$_{25}$ $\sim$ 4.5. Along
the minor axis, the exterior \HI ~is strongly asymmetric.
(5) In all four \HI ~maps, including that of NGC~4138, the velocity field 
is dominated by circular rotation,
but significant departures from motion in a quiescent disk are
evident. In all, the outer \HI ~velocity contours suggest significant warping of
the outer disk, roughly beginning at the optical edge and following Briggs' 
(1990) rule for warps, as commonly seen in other galaxies. 
The similiarities found in the five Sa galaxies
are consistent with the scenario that the kinematically distinct gas
components arise from slow, minor mergers of gas--rich satellites with an 
already--formed disk galaxy. Differences in their appearance thus
arise from differences in the circumstances of the merger events.
Retrograde primordial gas infall, as discussed
by Thakar \etal ~(1997) is
also offered as a possible solution to produce counter--rotating
components, although seemingly more difficult to
justify in the growing number of known cases. 

Though most numerical work on the merger process has focussed on equal mass
progenitors, recent studies have begun to address the minor merger
phenomenon (Quinn \etal ~1993; Hernquist \& Mihos 1995; Walker \etal ~1996; 
Thakar \etal ~1997;
Bekki 1998; Thakar \& Ryden 1998; Taniguchi ~1999). Although much of this
work is preliminary, requiring further refinement of the range of
galaxy characteristics and orbital parameters and the influence
of gas dissipation, triaxiality and dark matter content, we can nonetheless
examine the main characteristics of the five galaxies under the
assumption that we may be seeing the results of varying acquisition 
scenarios.

The extensive \HI ~distributions
seen in four of the five galaxies discussed here may result from the slow
accretion of a gas--rich satellite that is tidally stripped
before the merger occurs. Accretion of a low density gas-rich satellite
would minimize the disk heating problem and might 
favor the formation of extended \HI ~disks or rings, but perhaps not inner rings.
The formation of rings is seen in a variety of infalling gas
models, both due to orbit crowding and to true self--gravitating rings. Also,
Quinn \etal ~(1993) suggest that transport of angular
momentum outwards will tend to cause the disk to spread in radius. The newly--acquired
gas may reach sufficient densities, if clumpy, to form a new generation of stars.
This possibility is suggested by the faint, blue, low-surface brightness outer disk
seen in NGC~4772 and a similar excess of light
found around the counter-rotating Sab galaxy NGC~7217 (Buta \etal ~1995).

Clearly the timescales of the merger events responsible 
for the individual characteristics noted here are likely to vary greatly.
Relatively gas--poor and large--bulged, NGC~5854 may represent the
result of the accretion of a more massive satellite travelling in a prograde
orbit (Walker \etal ~1996). Merger events might lead to bulge build--up
and even induce large-scale star formation, leaving the galaxy in the 
post-starburst state in which NGC~5854 is found. Furthermore,
one of the primary arguments made against the tidal forcing of warps is that the
modes would damp out after a few galactic rotations. Perhaps in the cases
that suggest strong warping of the \HI, the minor mergers may be relatively 
recent events. Over time, the warps may damp, or be reexcited by additional
interactions. If there are no impediments to star formation in a counter-rotating
gas component, NGC~4138 with its two counter-rotating extended stellar
disks may represented the future evolutionary state of NGC~3626.

The similarities among the results for these five galaxies
lead to several questions and conundrums. 
(1) Is it key that the sample is composed of Sa galaxies?
The accretion of moderate amounts of gas may also sweep up the 
gas contained within the primary disk itself, ultimately leading
to a smoothing of the spiral structure.  
Do minor mergers lead to bulge build-up and disk heating, driving the
post-merger morphology towards earlier spiral types? 
(2) Damping the process of disk heating is a challenge for models of mergers, even
minor ones, if the accreted satellite is sufficiently massive or dense.
The resultant disk heating will be relatively lessened if the
satellite is predominantly gaseous. What are the conditions
necessary to avoid over-heating the disk and disturbing the overall
velocity field and spiral structure? 
(3) Is the fact that these galaxies are all unbarred important? 
Loss of angular momentum may drive the material inward toward the central regions
where it accumulates in a nuclear disk as is often discussed for the kinematically
decoupled cores seen in ellipticals. In the early stages of galaxy evolution,
the formation of bars seems inevitable, but if the central concentration of mass
is sufficient, the development of an inner Lindblad resonance will inhibit
the central flow of gas (Sellwood \& Moore 1999). Thus, since the progenitor primary
was already a well-developed object, infalling material may be stopped at the
inner Lindblad resonance, allowing the build--up of gas in a ring with subsequent
star formation.  Does this process produce the inner \Ha ~and \HI ~rings seen here?
Jungwiert \& Palous (1996) propose that unbarred ringed galaxies 
must possess weak central oval distortions, while the two--stream instability
picture of Lovelace \etal ~(1997) for galaxies with counter-rotating
components suggests the generation of strong m=1 spiral waves. Are the
inner kinematic
peculiarities evidence of such oval distortions or m=1 streaming?  
(4) Although these Sa galaxies are relatively isolated, they are all members
of loose groups. In fact, Sa galaxies are found typically
in denser environments than their later type counterparts. Is the relative
morphological segregation seen across the spiral sequence
related to the likelihood of minor mergers and
their increased probability in environments characterized by
moderate density and low velocity dispersion?

While examination of the rate of strongly disturbed galaxies may provide 
a reliable estimate of the major merger rate (e.g. Keel \& Wu, 1995), 
minor mergers may be morphologically hidden, revealed only by the presence 
of their resultant kinematic peculiarities. Because of the ambiguities
involved in disentangling the kinematic clues, it is not possible to
trace with any degree of precision the evolutionary history of the multiple stellar 
and gaseous components evident in these galaxies. Food for thought, however, is 
the possibility that, over the course of its lifetime, 
a disk galaxy might undergo multiple minor accretion events with kinematic 
as well as morphological consequences. In particular, more complicated histories
involving multiple minor mergers are likely to produce heterogenous
properties, particularly those related to properties of ``form''.
We suggest that the heterogeneity of the Sa class in particular
is the outcome of such complicated life histories, the memory of
which is signaled only by kinematic clues such as those identified here.
 
\acknowledgments

We thank Giovanni Moriondo for providing his 2--D bulge--decomposition code FITMAP
and Giovanni Moriondo, David E. Hogg and Morton S. Roberts for numerous 
discussions. This work has been supported by NSF grants AST--9023450 and 
AST--9528860 to MPH.
Portions of this research have been used in partial fulfillment of the
Ph.D. (KPJ) and M.Engr. (EAB) degrees at Cornell University.
BMM was supported by NSF-REU grant AST--9619531.
This study has made use of the NASA/IPAC Extragalactic
Database (NED) which is operated by the Jet Propulsion
Laboratory, Caltech, under contract with the National
Aeronautics and Space Administration and the Digital Sky Survey
which was produced at the Space
Telescope Science Institute under US Government grant NAG W-2166.

\vfil

\eject

\begin{figure}
\caption[]{
Cleaned continuum--subtracted \HI ~line channel maps of NGC~3626 derived from the
natural weight (low resolution) data cube.
Contour levels range from 0.46 to 4.6 mJy beam$^{-1}$ in increments
of 0.46 mJy beam$^{-1}$.  The beam size is 16\farcs8 x
16\farcs2. 
\label{fig1}}
\end{figure}

\begin{figure}
\caption[]{
Cleaned continuum--subtracted \HI ~line channel maps of NGC~3900 derived from the
natural weight (low resolution) data cube.
Contour levels range from 0.42 to 6.42 mJy beam$^{-1}$ in increments
of 1.0 mJy beam$^{-1}$.  The beam size is 22\farcs5 x
20\farcs0. 
\label{fig2}}
\end{figure}

\begin{figure}
\caption[]{
Cleaned continuum--subtracted \HI ~line channel maps of NGC~4772 derived from the
natural weight (low resolution) data cube.
Contour levels range from 0.62 to 8.62 mJy beam$^{-1}$ in increments
of 1.0 mJy beam$^{-1}$.  The beam size is 31\farcs7 x 23\farcs1.
\label{fig3}}
\end{figure}

\begin{figure}
\caption[]{
Global profiles derived from VLA mapping for the target galaxies
and neighboring companions. \HI ~parameters are given in Table 8.
\label{fig4}}
\end{figure}

\begin{figure}
\caption[]{
Imaging results for NGC~3626. Top left panel: $R$ band image with 
the scaling set to showing the maximum extent of starlight. Middle left panel:
Expanded $R$ band view illustrating the faint ring and spiral structure. 
The ellipse has the dimensions of D$_{25}$ $\times$ d$_{25}$ and
is oriented along the optical major axis. Lower left panel:
The inner 1\arcmin~ shown to emphasize the dustlane west of the nucleus. 
Upper right panel: Results of the two--dimensional bulge--disk decomposition
for a $n=2$ bulge. The solid line traces the light distribution while the
disk and bulge are shown as long--dashed and short--dashed lines
respectively. 
Middle right panel: The high resolution \HI ~column
density map superposed on a greyscale representation of the central $R$ band image
The first \HI ~contour is at
1.0$\times 10^{20}$ atoms cm$^{-2}$.  The next six contours range from 4.0 to 
11.5$\times 10^{20}$ atoms cm$^{-2}$ in steps of 1.5$\times 10^{20}$ atoms cm$^{-2}$
and the remainder range from 13.5 to 18.5$\times 10^{20}$ atoms cm$^{-2}$
in steps of 1.0$\times 10^{20}$ atoms cm$^{-2}$. Lower right panel:
Identical contours of the high resolution \HI ~column
density map superposed on the \Ha ~image.
\label{Haynes.fig5}}
\end{figure}

\begin{figure}
\caption[]{
\HI ~mapping results obtained
from the low resolution cube for NGC~3626 and UGC~6341.
Left panel: \HI ~column density contours and grayscale ranging from 1.0 
(2$\sigma$) to 17.5$\times 10^{20}$ atoms cm$^{-2}$ in steps of 
1.5$\times 10^{20}$ atoms cm$^{-2}$. Middle panel: The same
\HI ~column density map overlayed on the $R$ band image.  The optical
greyscale has the same shading as in the center left panel of Figure 5.
Right panel: The \HI ~velocity field map. Velocity contours range from 
1275 (light) to 1675 (dark) \kms~ in increments of 25 \kms. 
In all three panels, the ellipse outlines the optical galaxy as given
by $D_{25} \times d_{25}$ oriented along the optical major axis,
PA = 157\arcdeg.
\label{fig6}}
\end{figure}

\begin{figure}
\caption[]{
Position--velocity cuts obtained from the low resolution \HI ~cube
and the optical spectra for NGC~3626.
Upper panels: \HI ~position--velocity diagrams along four slices through the low
resolution data cube.
The top left panel shows the slice along the optical
major axis, PA = 157\arcdeg, along which the optical spectra were obtained.
In addition to the \HI ~contours, the velocities of the optical
stellar absorption line (open squares) and emission line (filled
circles) are also shown. The secondary gas component is represented
by open circles. The top right panel shows the slice at PA = 178\arcdeg,
the major axis of the outer \HI ~distribution. The center left panel
shows the slice along the optical
minor axis, PA = 67\arcdeg; the velocities of the stellar and ionized
gas components obtained from the long slit spectra are also shown.
The center right panel shows the slice along a position angle
of 90\arcdeg, the minor axis of the outer \HI ~ring where the
velocity gradient is minimized. The lower panels show
the optical rotation curves derived for NGC~3626 along the major
axis (PA = 157\arcdeg) and along the minor axis (PA = 67\arcdeg). Separate symbols
show the rotation curves derived from the stellar absorption and gas
components. Error bars are shown.
In the innermost regions, the [NII] lines appear double.
\label{fig7}}
\end{figure}

\begin{figure}
\caption[]{
Results of optical imaging for NGC~3900. Upper panel: 
$B$ band image with scale set to emphasize outer
structure. The ellipse has the dimensions of D$_{25}$ $\times$ d$_{25}$ and
is oriented along the optical major axis, PA = 2\arcdeg. 
Middle panel: Low contrast $B$ band grayscale showing spiral structure
and dustlanes in the inner galaxy. The spatial scale is zoomed by
a factor of two relative to the upper panel. Lower panel: 
Results of the two--dimensional bulge--disk decomposition
for a $n=2$ bulge. The solid line traces the light distribution while the
disk and bulge are shown as long--dashed and short--dashed lines
respectively. 
\label{fig:fig8}}
\end{figure}

\begin{figure}
\caption[]{
\HI ~mapping results  obtained
from the low resolution cube for NGC~3900.
Left panel: \HI ~column density contours ranging from  
0.3 (2$\sigma$) to 12.3$\times 10^{20}$ atoms cm$^{-2}$ in steps of 
2.0$\times 10^{20}$ atoms cm$^{-2}$. Middle panel: Similar \HI ~column 
density map, with a stepsize of 1.0$\times 10^{20}$ atoms cm$^{-2}$
overlayed on the $B$ band image. Right panel: The \HI ~velocity 
field map showing velocity contours ranging from 1575 (light) to 2050 (dark)
\kms ~in increments of 25 \kms. 
In all three panels, the ellipse outlines the optical galaxy as given
by $D_{25} \times d_{25}$ oriented along the optical major
axis PA = 2\arcdeg.
\label{fig:fig9}}
\end{figure}

\begin{figure}
\caption[]{
Position--velocity cuts obtained from the low resolution \HI ~cube
and the optical spectra for NGC~3900.
Top panels: \HI ~position--velocity diagrams along two slices through the low
resolution data cube.
The top left panel shows the slice along the optical
major axis, PA = 2\arcdeg, along which the optical spectra were obtained.
In addition to the \HI ~contours, the rotation curves traced by the optical
stellar absorption lines (open squares) and the gas emission lines (filled
circles) are also shown.  The top right panel shows the slice at PA = 92\arcdeg,
along the minor axis. The lower panels show
the optical rotation curves derived for NGC~3900 along the major
(PA = 2\arcdeg) and minor (PA = 92\arcdeg) axes. Separate symbols
show the rotation curves derived from the stellar absorption and gas
components. Error bars are shown. 
\label{fig10}}
\end{figure}

\begin{figure}
\caption[]{
Imaging results for NGC~4772. Top left panel: $B$ band
image scaled to show the faint outer disk. Middle left panel:
B-band image of the same region but with the grayscale set
to emphasize the inner regions, showing that the dustlane cuts
across the bulge in the NE. Lower left panel: \Ha ~image.
The ellipse superposed on the three images has the dimensions of 
D$_{25}$ $\times$ d$_{25}$ and is oriented along the optical major axis,
PA = 147\arcdeg. 
Upper right panel: Results of the two--dimensional bulge--disk decomposition
for a $n=2$ bulge. The solid line traces the light distribution while the
disk and bulge are shown as long--dashed and short--dashed lines
respectively. Middle right panel: The high resolution \HI ~column
density map superposed on a greyscale representation of the $B$ band image
in the inner regions.
Contours extend from 0.6$\times 10^{20}$ to 14$\times 10^{20}$ 
atoms cm$^{-2}$ in steps of 4.0$\times 10^{20}$ atoms cm$^{-2}$.
Lower right panel: Identical contours of the high resolution \HI ~column
density map superposed on the \Ha ~image.
\label{fig11}}
\end{figure}

\begin{figure}
\caption[]{
Results for the large-scale \HI ~mapping of NGC~4772. Upper left panel:
\HI ~column density contours superposed on the $B$ band image. Contours are shown
at levels of 0.5, 1.5, 2.5, 4.0, 6.0, 8.0, 10.0 and 12.0 and 14.0 $\times 
10^{20}$ atoms cm$^{-2}$. Upper left panel: Greyscale and contours of
the \HI ~column density as shown in the left panel. Lower left panel:
The outermost HI
column density contour at 5$\times 10^{19}$ atoms cm$^{-2}$ superposed on a
grayscale representation of the $B$ band image scaled to enhance the
faint outer disk.  Lower right panel: The velocity field of NGC~4772,
shown as contours and grayscale extending from 
770 \kms~ (NW) to 1250 \kms~  (SE) in steps of 30 \kms. 
In all panels, the ellipse has dimensions D$_{25}$ $\times$ d$_{25}$ and
is oriented along the optical major axis, PA = 147\arcdeg.
\label{fig12}}
\end{figure}

\begin{figure}
\caption[]{
Position--velocity cuts obtained from the low resolution \HI ~cube
and the optical spectra for NGC~4772.
Top panels: \HI ~position--velocity diagrams along two slices through the low
resolution data cube. 
The top right panel shows the slice along the optical
major axis, PA = 147\arcdeg, along which the optical spectra were obtained.
In addition to the \HI ~contours, the velocities derived from the optical
stellar absorption lines (open squares) and ionized gas emission lines (filled
circles) are also shown.  The top right panel shows the slice at PA = 57\arcdeg,
along the minor axis. The lower panels show
the optical rotation curves derived for NGC~4772 along the major
axis (PA = 147\arcdeg) and along the minor axis (PA = 57\arcdeg). Separate symbols
show the rotation curves derived from the stellar absorption and gas
components. Error bars are shown. 
\label{fig13}}
\end{figure}

\begin{figure}
\caption[]{
Results for NGC~5854. Upper left panel: $R$ band image. The ellipse has dimensions 
D$_{25}$ $\times$ d$_{25}$ and is oriented along the optical major axis, PA = 55\arcdeg.
Upper right panel: Results of the two--dimensional bulge--disk decomposition
for a $n=4$ bulge. The solid line traces the light distribution while the
disk and bulge are shown as long--dashed and short--dashed lines
respectively. Center right panel: Central channels of the \HI ~single dish
spectrum obtained at Arecibo by Magri (1994).  The lower panels show
the optical rotation curves derived for NGC~5854 along the major
axis (PA = 55\arcdeg) and along the minor axis (PA = 145\arcdeg). Separate symbols
show the rotation curves derived from the stellar absorption and gas
components. Error bars are shown. 
\label{fig14}}
\end{figure}

\begin{figure}
\caption[]{
New imaging results for NGC~4138. Top left panel: $B$ band image scaled to illustrate
the inner bulge, disk and dustlane. The ellipse has dimensions 
D$_{25}$ $\times$ d$_{25}$ and is oriented along the optical major axis,
PA = 150\arcdeg.
Top right panel: Results of the two--dimensional bulge--disk decomposition
for a $n=2$ bulge. The solid line traces the light distribution while the
disk and bulge are shown as long--dashed and short--dashed lines
respectively. Center left panel: The inner \HI ~contours
superposed on a zoomed $B$ band image. The contours are at levels of
20 to 95 $\times 10^{19}$ in 
steps of 15 $\times 10^{19}$ atoms cm$^{-2}$. Center right panel: Similar 
display of the \HI ~superposed on the H--alpha image.  
Bottom left panel: \HI ~column density contours superposed on the $V$ band
image. The \HI ~map shows
contours at levels of 3.4, 6.9 and 13.8 $\times 10^{19}$ atoms cm$^{-2}$ plus contours 
from 20 to 95 $\times 10^{19}$ atoms cm$^{-2}$ in steps of 15 atoms cm$^{-2}$.
For further details, see Jore \etal ~(1996).
The ellipse is the same as in the upper left panel. Bottom right: Identical
\HI ~column density contours as the bottom left panel, here
also shown as grayscale to emphasize the 
patchiness of the outer \HI ~distribution.
\label{fig15}}
\end{figure}
\end{document}